\documentclass[a4paper,UKenglish]{lipics-v2016}

\edef\keptrmdefault{\rmdefault}
\edef\keptsfdefault{\sfdefault}
\edef\keptttdefault{\ttdefault}
\usepackage[math]{iwona}
\edef\rmdefault{\keptrmdefault}
\edef\sfdefault{\keptsfdefault}
\edef\ttdefault{\keptttdefault}

\usepackage[usenames]{xcolor}
\usepackage{mathpartir}
\usepackage{xspace}
\usepackage{mathtools}
\usepackage{breakurl}
\usepackage[scaled=0.75]{beramono}
\usepackage{tikz}
\usepackage{adjustbox}

\newcommand*{\piCalculus}{$\pi$-calculus\xspace}

\newcommand*{\etal}{et al\xspace}

\newcommand*{\ttt}[1]{\texttt{#1}}

\newcommand*{\eqdef}{\stackrel{\smash{\text{\tiny def}}}{=}}

\newcommand*{\Cdot}{\raisebox{-0.25ex}{\scalebox{1.4}{$\cdot$}}}
\newcommand*{\set}[1]{\{#1\}}

\newcommand*{\join}{\sqcup}
\newcommand*{\meet}{\sqcap}

\newcommand*{\lowlight}[1]{\textcolor{darkgray}{#1}}
\newcommand*{\sub}[2]{#1_{\lowlight{#2}}}

\newcommand*{\param}{\cdot}

\newcommand*{\appref}[1]{Appendix~\ref{app:#1}}

\newcommand*{\defref}[1]{Definition~\ref{def:#1}}

\newcommand*{\figref}[1]{Figure~\ref{fig:#1}}
\newcommand*{\figrefTwo}[2]{Figures \ref{fig:#1} and \ref{fig:#2}}
\newcommand*{\lemref}[1]{Lemma~\ref{lem:#1}}

\newcommand*{\secref}[1]{\S\,\ref{sec:#1}}

\newcommand*{\thmref}[1]{Theorem~\ref{thm:#1}}

\newenvironment{nop}{}{}
\newenvironment{sdisplaymath}{
\begin{nop}\small\begin{displaymath}}{
\end{displaymath}\end{nop}\ignorespacesafterend}
\newenvironment{smathpar}{
\begin{nop}\small\begin{mathpar}}{
\end{mathpar}\end{nop}\ignorespacesafterend}

\newenvironment{mathfig}{\begin{sdisplaymath}}{\end{sdisplaymath}}
\newenvironment{syntaxfig}{\begin{mathfig}\begin{array}{@{}l@{\quad}r@{~~}c@{\quad}ll}}{\end{array}\end{mathfig}}

\newenvironment{nscenter}
 {\parskip=0pt\par\nopagebreak\centering}
 {\par\noindent\ignorespacesafterend}

\DeclareFontFamily{U}{MnSymbolC}{}
\DeclareFontShape{U}{MnSymbolC}{m}{n}{
    <-6>  MnSymbolC5
   <6-7>  MnSymbolC6
   <7-8>  MnSymbolC7
   <8-9>  MnSymbolC8
   <9-10> MnSymbolC9
  <10-12> MnSymbolC10
  <12->   MnSymbolC12}{}
\DeclareFontShape{U}{MnSymbolC}{b}{n}{
    <-6>  MnSymbolC-Bold5
   <6-7>  MnSymbolC-Bold6
   <7-8>  MnSymbolC-Bold7
   <8-9>  MnSymbolC-Bold8
   <9-10> MnSymbolC-Bold9
  <10-12> MnSymbolC-Bold10
  <12->   MnSymbolC-Bold12}{}
\DeclareSymbolFont{MnSyC}{U}{MnSymbolC}{m}{n}

\DeclareMathSymbol{\medsquare}{\mathord}{MnSyC}{106}

\DeclareFontFamily{U}{MnSymbolD}{}
\DeclareFontShape{U}{MnSymbolD}{m}{n}{
    <-6>  MnSymbolD5
   <6-7>  MnSymbolD6
   <7-8>  MnSymbolD7
   <8-9>  MnSymbolD8
   <9-10> MnSymbolD9
  <10-12> MnSymbolD10
  <12->   MnSymbolD12}{}
\DeclareFontShape{U}{MnSymbolD}{b}{n}{
    <-6>  MnSymbolD-Bold5
   <6-7>  MnSymbolD-Bold6
   <7-8>  MnSymbolD-Bold7
   <8-9>  MnSymbolD-Bold8
   <9-10> MnSymbolD-Bold9
  <10-12> MnSymbolD-Bold10
  <12->   MnSymbolD-Bold12}{}
\DeclareSymbolFont{MnSyD}{U}{MnSymbolD}{m}{n}

\DeclareMathSymbol{\Cong}{\mathrel}{MnSyD}{12}

\DeclareFontFamily{U}{MnSymbolC}{}
\DeclareFontShape{U}{MnSymbolC}{m}{n}{
    <-6>  MnSymbolC5
   <6-7>  MnSymbolC6
   <7-8>  MnSymbolC7
   <8-9>  MnSymbolC8
   <9-10> MnSymbolC9
  <10-12> MnSymbolC10
  <12->   MnSymbolC12}{}
\DeclareFontShape{U}{MnSymbolC}{b}{n}{
    <-6>  MnSymbolC-Bold5
   <6-7>  MnSymbolC-Bold6
   <7-8>  MnSymbolC-Bold7
   <8-9>  MnSymbolC-Bold8
   <9-10> MnSymbolC-Bold9
  <10-12> MnSymbolC-Bold10
  <12->   MnSymbolC-Bold12}{}
\DeclareSymbolFont{MnSyC}{U}{MnSymbolC}{m}{n}

\DeclareMathSymbol{\MnSymbolvdots}{\mathord}{MnSyC}{6}

\DeclareFontFamily{U}{matha}{\hyphenchar\font45}
\DeclareFontShape{U}{matha}{m}{n}{
      <5> <6> <7> <8> <9> <10> gen * matha
      <10.95> matha10 <12> <14.4> <17.28> <20.74> <24.88> matha12
      }{}
\DeclareSymbolFont{matha}{U}{matha}{m}{n}

\DeclareMathSymbol{\twoPrime}{3}{matha}{"32}
\DeclareMathSymbol{\threePrime}{3}{matha}{"33}
\DeclareMathSymbol{\fourPrime}{3}{matha}{"34}

\everymath{\displaystyle}

\usepackage{microtype}

\hypersetup{ 
    colorlinks=false,
    pdfborder={0 0 0},
}

\newcommand{\crossrule}{\noindent\textcolor{lightgray}{\cleaders\hbox{.}\hfill}}

\makeatletter
\@fleqnfalse
\@mathmargin\@centering
\makeatother

\newcommand*{\piBoundOutput}[1]{\compl{#1}}
\newcommand*{\piBoundOutputN}[2]{\compl{#1}(#2)}
\newcommand*{\piInput}[1]{\underline{#1}}
\newcommand*{\piInputN}[2]{#1(#2)}
\newcommand*{\piOutput}[2]{\compl{#1}\langle#2\rangle}
\newcommand*{\piTau}{\tau}

\newcommand*{\piAction}[2]{#1.#2}
\newcommand*{\piChoice}[2]{{#1} + {#2}}

\newcommand*{\piPar}[2]{{#1} \mid {#2}}

\newcommand*{\piReplicate}[1]{{!#1}}
\newcommand*{\piRestrictN}[2]{(\nu#1)\;#2}
\newcommand*{\piRestrict}[1]{\nu#1}

\newcommand*{\piZero}[0]{\mathbf{0}}

\newcommand*{\targetF}{\textsf{tgt}}

\newcommand*{\target}[1]{\targetF({#1})}

\newlength{\arrowlen}
\newcommand*{\myrightarrow}[1]{\xrightarrow{\mathmakebox[\arrowlen]{#1}}}
\newcommand*{\myrightharpoondown}[1]{\xrightharpoondown{\mathmakebox[\arrowlen]{#1}}}
\newcommand*{\myleftharpoondown}[1]{\xleftharpoondown{\mathmakebox[\arrowlen]{#1}}}
\newcommand*{\transition}[1]{\myrightarrow{\smash{#1}}}
\newcommand*{\transitionWithoutSmash}[1]{\myrightarrow{#1}}
\newcommand*{\compl}[1]{\overline{#1}}

\newcommand*{\concur}{\smile}
\newcommand*{\residual}[2]{{#1}\slash{#2}}
\newcommand*{\plus}[2]{#1 + #2}
\newcommand*{\suc}[1]{\plus{#1}{1}}
\newcommand*{\ren}[2]{\renRaw{#1} {#2}}
\newcommand*{\renRaw}[1]{{#1}^*}
\newcommand*{\id}{\textsf{id}}
\newcommand*{\swap}[1]{\sub{\swapR}{#1}}
\newcommand*{\swapR}{\textsf{swap}}
\newcommand*{\pop}[2]{\sub{\popR}{{#1}}\;{#2}}
\newcommand*{\popR}{\textsf{pop}}
\newcommand*{\push}[1]{\sub{\pushR}{#1}}
\newcommand*{\pushR}{\textsf{push}}

\newcommand*{\Action}[1]{\textsf{Action}\;{#1}}

\newcommand*{\congRestrictSwap}[1]{\sub{\nu\nu\text{-}\swapR}{#1}} 
\newcommand*{\freeBraid}{\mathrel{\protect\rotatebox[origin=c]{90}{$\ltimes$}}}
\newcommand*{\boundBraid}{\mathrel{\protect\rotatebox[origin=c]{90}{$\rtimes$}}}

\newcommand*{\permEq}{\simeq}
\newcommand*{\cons}{\mathrel{\Cdot}}
\newcommand*{\nil}{\varepsilon}
\newcommand*{\transitionVec}[1]{\mathrel{\myrightarrow{#1}^*}}
\newcommand*{\braiding}[1]{\sub{\gamma}{#1}}

\renewcommand*{\vec}[1]{\bm{#1}}
\newcommand*{\after}{\circ}
\renewcommand*{\to}{\longrightarrow}
\newcommand*{\hole}{\scalebox{0.8}{$\medsquare$}}

\newcommand{\rtransition}[1]{\myleftharpoondown{\smash{#1}}}
\newcommand{\ftransition}[1]{\myrightharpoondown{\smash{#1}}}

\newcommand{\unrenF}[1]{\sub{\textsf{unren}}{#1}}
\newcommand{\renF}[1]{\sub{\textsf{ren}}{#1}}
\newcommand{\renAppl}[2]{\renApplF{#1}(#2)}
\newcommand{\renApplF}[1]{\sub{\textsf{app}}{#1}}

\newcommand{\renUnapplF}[1]{\sub{\textsf{unapp}}{#1}}
\newcommand{\renMapsTo}[3]{#2\sub{\mapsto}{#1}#3}
\newcommand{\renUnapp}[3]{\renUnappF{#1}{#2}#3}
\newcommand{\renUnappF}[2]{#1^{\scriptscriptstyle -1}_{#2}}
\newcommand{\stepF}[1]{\sub{\textsf{step}}{#1}}
\newcommand{\step}[2]{\stepF{#1}\;#2}
\newcommand{\stepGC}[1]{\sub{\overline{\textsf{step}}}{#1}}
\newcommand{\unstepF}[1]{\sub{\textsf{unstep}}{#1}}
\newcommand{\unstep}[2]{\unstepF{#1}\;#2}
\newcommand{\stepNoActionF}[1]{\sub{\textsf{step}'}{#1}}
\newcommand{\stepNoAction}[2]{\stepNoActionF{#1}\;#2}

\newcommand{\unstepNoActionF}[1]{\sub{\textsf{unstep}'}{#1}}
\newcommand{\unstepNoAction}[2]{\unstepNoActionF{#1}\;#2}
\newcommand{\fwdF}[1]{\sub{\textsf{fwd}}{#1}}
\newcommand{\fwd}[2]{\fwdF{#1}\;#2}

\newcommand{\bwdF}[1]{\sub{\textsf{bwd}}{#1}}
\newcommand{\bwd}[2]{\bwdF{#1}\;#2}
\newcommand{\mapF}[1]{\sub{\textsf{map}}{#1}}

\newcommand{\mapGC}[1]{\sub{\overline{\textsf{map}}}{#1}}
\newcommand{\unmapF}[1]{\sub{\textsf{unmap}}{#1}}

\newcommand{\braidGC}[1]{\sub{\overline{\textsf{braid}}}{#1}}
\newcommand{\braidF}[1]{\sub{\textsf{braid}}{#1}}
\newcommand{\braid}[2]{\braidF{#1}\;#2}
\newcommand{\unbraidF}[1]{\sub{\textsf{unbraid}}{#1}}
\newcommand{\unbraid}[2]{\unbraidF{#1}\;#2}

\renewcommand*{\vec}[1]{\tilde{#1}}

\newcommand{\subst}[3]{#1\{#2/#3\}}
\newcommand{\redex}[1]{\mathbf{#1}}
\newcommand{\slice}[1]{\textcolor{lightgray}{#1}}

\newcommand{\down}[1]{{\mathrel{\raisebox{0.08em}{$\downarrow$}}}#1}

\newcommand{\Paragraph}[1]{\textbf{#1}.}
\newcommand{\ConcurrentSlicing}{\ttt{concurrent-slicing}\xspace}

\EventEditors{Jos\'ee Desharnais and Radha Jagadeesan}
\EventNoEds{2}
\EventLongTitle{27th International Conference on Concurrency Theory (CONCUR 2016)}
\EventShortTitle{CONCUR 2016} \EventAcronym{CONCUR}
\EventYear{2016}
\EventDate{August 24--27, 2016} \EventLocation{Qu\'ebec City, Canada}
\EventLogo{}
\SeriesVolume{59}
\ArticleNo{18}

\begin{document}

\title{Causally consistent dynamic slicing}
\author[1,2]{Roly Perera}
\author[3]{Deepak Garg}
\author[1]{James Cheney}
\affil[1]{Laboratory for Foundations of Computer Science, University of Edinburgh, Edinburgh, UK\\
  \texttt{rperera@inf.ed.ac.uk}, \texttt{jcheney@inf.ed.ac.uk}}
\affil[2]{School of Computing Science, University of Glasgow, Glasgow, UK\\
  \texttt{roly.perera@glasgow.ac.uk}}
\affil[3]{Max Planck Institute for Software Systems, Saarbr\"ucken, Germany\\
  \texttt{dg@mpi-sws.org}}
\authorrunning{R. Perera, D. Garg and J. Cheney}

\Copyright{Roly Perera, Deepak Garg and James Cheney}

\subjclass{D.1.3 Concurrent Programming; D.2.5 Testing and debugging}
\keywords{\piCalculus; dynamic slicing; causal equivalence; Galois connection}
\maketitle

\begin{abstract}
We offer a lattice-theoretic account of dynamic slicing for \piCalculus,
building on prior work in the sequential setting. For any run of a
concurrent program, we exhibit a Galois connection relating forward
slices of the start configuration to backward slices of the end
configuration. We prove that, up to lattice isomorphism, the same Galois
connection arises for any causally equivalent execution, allowing an
efficient concurrent implementation of slicing via a standard
interleaving semantics. Our approach has been formalised in the
dependently-typed language Agda.
\end{abstract}

\section{Introduction}
\label{sec:introduction}

Dynamic slicing, due originally to Weiser \cite{weiser81}, is a runtime
analysis technique with applications in debugging, security and
provenance tracking. The basic goal is to identify a sub-program, or
\emph{program slice}, that may affect an outcome of interest called the
\emph{slicing criterion}, such as the value of a variable. Dynamic
slicing in concurrent settings is often represented as a graph
reachability problem, thanks to influential work by Cheng
\cite{cheng93}. However, most prior work on dynamic slicing for
concurrency does not yield minimum slices, nor allows particularly
flexible slicing criteria, such as arbitrary parts of configurations.
Systems work on concurrent slicing \cite{goswami00,manandhar04,tallam08}
tends to be largely informal.

Perera \etal \cite{perera12a} developed an approach where backward
dynamic slicing is treated as a kind of (abstract) reverse execution or
``rewind'' and forward slicing as a kind of (abstract) re-execution or
``replay''. Forward and backward slices are related by a Galois
connection, ensuring the existence of minimal slices. This idea is
straightforward in the sequential setting of the earlier work. However,
generalising it to concurrent programs is non-trivial. Suppose we run a
concurrent computation, discover a bug, and then wish to compute a
dynamic slice. It would clearly be impractical to require the slice be
computed using the exact interleaving of the original run, particularly
in a distributed setting. On the other hand, computing the slice using a
brand-new concurrent execution may make different non-deterministic
choices, producing a slice of a computation other than the one intended.

Intuitively, any execution which exhibits the same causal structure
should be adequate for computing the slice, and any practical approach
to concurrent slicing should take advantage of this. Danos and Krivine
\cite{danos04a} make a similar observation about reversible concurrency,
arguing that the most liberal notion of reversibility is one that just
respects causality: an action can only be undone after all the actions
that causally depend on it have been undone.

In this paper we formalise dynamic slicing for \piCalculus, and show
that any causally equivalent execution generates precisely the same
slicing information. We do this by formalising slicing with respect to a
particular execution $\vec{t}$, and then proving that slicing with
respect to any causally equivalent computation $\vec{u}$ yields the same
slice, after a unique ``rewiring'' which interprets the path witnessing
$\vec{t} \permEq \vec{u}$ as a lattice isomorphism relating the two
slices. The isomorphism is constructive, rewriting one slice into the
other: this allows non-deterministic metadata (e.g.~memory addresses or
transaction ids) in the slicing execution to be aligned with the
corresponding metadata in the original run. We build on an earlier
``proof-relevant'' formalisation of causal equivalence for \piCalculus
in Agda \cite{perera16}. As long as causality is respected, an
implementation of our system can safely use any technique (e.g.~redex
trails, proved transitions, or thread-local memories) to implement
rewind and replay.

\Paragraph{Example: scheduler with non-compliant task}
While dynamic slicing cannot automatically isolate bugs, it can hide
irrelevant detail and yield compact provenance-like explanations of
troublesome parts of configurations. As an example we consider Milner's
scheduler implementation \cite[p.~65]{milner99}. The scheduler controls
a set of $n$ tasks, executed by agents $A_1, \ldots, A_n$. Agent $A_i$
sends the message $a_i$ (\emph{announce}) to the scheduler to start its
task, and message $b_i$ (\emph{break}) to end its task. The scheduler
ensures that the actions $a_i$ occur cyclically starting with $a_1$, and
that for each $i$ the actions $a_i$ and $b_i$ alternate, starting with
$a_i$. Although started sequentially, once started the tasks are free to
execute in parallel.

\begin{figure}
\begin{nscenter}
{\small
\[\begin{array}{llllllll}
&
\textit{Scheduler thread 1}
&&
\textit{Scheduler thread 2}
&&
A_1
&&
A_2
\\[1mm]
&
\redex{a_1.}c_1.(b_1.\compl{c_2}.\compl{r_1} + \slice{\compl{c_2}.b_1.\compl{r_1}})
& \mid &
\compl{c_1}.a_2.c_2.\slice{(b_2.\compl{c_1}.\compl{r_2} + \compl{c_1}.b_2.\compl{r_2})}
& \mid &
\redex{\compl{a_1}.}\slice{\compl{b_1}.\compl{p_1}}
& \mid &
\compl{a_2}.\compl{b_1}.\slice{\compl{p_2}}
\\
\transition{} &
\redex{c_1.}(b_1.\compl{c_2}.\compl{r_1} + \slice{\compl{c_2}.b_1.\compl{r_1}})
& \mid &
\redex{\compl{c_1}.}a_2.c_2.\slice{(b_2.\compl{c_1}.\compl{r_2} + \compl{c_1}.b_2.\compl{r_2})}
& \slice{\mid} &
\slice{\compl{b_1}.\compl{p_1}}
& \mid &
\compl{a_2}.\compl{b_1}.\slice{\compl{p_2}}
\\
\transition{} &
b_1.\compl{c_2}.\compl{r_1} + \slice{\compl{c_2}.b_1.\compl{r_1}}
& \mid &
\redex{a_2.}c_2.\slice{(b_2.\compl{c_1}.\compl{r_2} + \compl{c_1}.b_2.\compl{r_2})}
& \slice{\mid} &
\slice{\compl{b_1}.\compl{p_1}}
& \mid &
\redex{\compl{a_2}.}\compl{b_1}.\slice{\compl{p_2}}
\\
\transition{} &
\redex{b_1.}\compl{c_2}.\compl{r_1} + \slice{\compl{c_2}.b_1.\compl{r_1}}
& \mid &
c_2.\slice{(b_2.\compl{c_1}.\compl{r_2} + \compl{c_1}.b_2.\compl{r_2})}
& \slice{\mid} &
\slice{\compl{b_1}.\compl{p_1}}
& \mid &
\redex{\compl{b_1}.}\slice{\compl{p_2}}
\\
\transition{} &
\redex{\compl{c_2}.}\compl{r_1}
& \mid &
\redex{c_2.}\slice{(b_2.\compl{c_1}.\compl{r_2} + \compl{c_1}.b_2.\compl{r_2})}
& \slice{\mid} &
\slice{\compl{b_1}.\compl{p_1}}
& \slice{\mid} &
\slice{\compl{a_2}.\compl{b_1}.\compl{p_2}}
\\
\transition{} &
a_1.\slice{c_1.(b_1.\compl{c_2}.\compl{r_1} + \compl{c_2}.b_1.\compl{r_1})}
& \slice{\mid} &
\slice{b_2.\compl{c_1}.\compl{r_2} + \compl{c_1}.b_2.\compl{r_2}}
& \slice{\mid} &
\slice{\compl{b_1}.\compl{p_1}}
& \slice{\mid} &
\slice{\compl{a_2}.\compl{b_1}.\compl{p_2}}
\end{array}\]}
\crossrule
\end{nscenter}
\caption{Stuck configuration, overlaid with backward slice with respect
  to final state of thread 1}
\label{fig:example:scheduler-run}
\end{figure}

\figref{example:scheduler-run} shows five transitions of a two-thread
scheduler, with the redex selected at each step highlighted in bold. The
parts of the configuration which contribute to the final state of thread
1 are in black; the grey parts are discarded by our backward-slicing
algorithm. Assume prefixing binds more tightly than either
$\piPar{\param}{\param}$ or $+$. To save space, we omit the
$\nu$-binders defining the various names, and write
$\piAction{x}{\piZero}$ simply as $x$. The names $r_1$, $r_2$, $p_1$ and
$p_2$ are used to make recursive calls \cite[p.~94]{milner99}: a
recursive procedure is implemented as a server which waits for an
invocation request, spawns a new copy of the procedure body, and then
returns to the wait state. Here we omit the server definitions, and
simply replace a successful invocation by the spawned body; thus in the
final step of \figref{example:scheduler-run}, after the synchronisation
on $c_2$ the invocation $\compl{r_1}$ is replaced by a fresh copy of the
initial state of scheduler thread 1.

The final state of \figref{example:scheduler-run} has no redexes, and so
is stuck. The slice helps highlight the fact that by the time we come to
start the second loop of scheduler 1, the task was terminated by message
$\compl{b_1}$ from $A_2$, before any such message could be sent by
$A_1$. We can understand the slice of the initial configuration
(computed by ``rewinding'', or backward-slicing) as \emph{sufficient} to
explain the slice of the stuck configuration by noting that the former
is able to \emph{compute} the latter by ``replay'', or forward-slicing.
In other words, writing a sliced part of the configuration as $\hole$,
and pretending the holes $\hole$ are sub-computations which get stuck,
we can derive

\vspace{-10pt}
{\small\[
\redex{a_1.}c_1.(b_1.\hole + \hole)
\mid
\compl{c_1}.a_2.\hole
\mid
\redex{\compl{a_1}.}\hole
\mid
\compl{a_2}.\compl{b_1}.\compl{p_2}
\quad
\transitionVec{}
\quad
\compl{a_2}.\hole
\]}

\vspace{-10pt}
\noindent without getting stuck. The slice on the left may of course
choose to take the right-hand branch of the choice instead. But if we
constrain the replay of the sliced program to follow the causal
structure of the original unsliced run -- to take the same branches of
internal choices, and have the same synchronisation structure -- then it
will indeed evolve to the slice on the right. This illustrates the
correctness property for backward slicing, which is that forward-slicing
its result must recompute (at least) the slicing criterion.

For this example, the tasks are entirely atomic and so fixing the
outcome of $+$ has the effect of making the computation completely
sequential. Less trivial systems usually have multiple ways they can
evolve, even once the causal structure is fixed. A confluence lemma
typically formalises the observational equivalence of two causally
equivalent runs. However, a key observation made in \cite{perera16} is
that requiring causally equivalent runs to reach exactly the same state
is too restrictive for \piCalculus, in particular because of name
extrusion. As we discuss in \secref{slicing-concurrent}, two causally
unrelated extrusion-rendezvous lead to states which differ in the
relative position of two $\nu$-binders, reflecting the two possible
orderings of the rendezvous. Although technically unobservable to the
program, interleaving-sensitive metadata, such as memory locations in a
debugger or transaction ids in a financial application, may be important
for domain-specific reasons. In these situations being able to robustly
translate between the target states of the two executions may be useful.

\Paragraph{Summary of contributions}
\secref{slicing} defines the core forward and backward dynamic slicing
operations for \piCalculus transitions and sequences of transitions
(traces). We prove that they are related by a Galois connection, showing
that backward and forward slicing, as defined, are minimal and maximal
with respect to each other. \secref{slicing-concurrent} extends this
framework to show that the Galois connections for causally equivalent
traces compute the same slices up to lattice isomorphism.
\secref{related-work} discusses related work and \secref{conclusion}
offers closing thoughts and prospects for follow-up work.
\appref{module-structure} summarises the Agda module structure and
required libraries; the source code can be found at
\url{https://github.com/rolyp/concurrent-slicing}, release \ttt{0.1}.

\section{Galois connections for slicing \piCalculus programs}
\label{sec:slicing}

To summarise informally, our approach is to interpret, functorially,
every transition diagram in the \piCalculus into the category of
lattices and Galois connections. For example the interpretation of the
transition diagram on the left is the commutative diagram on the right:

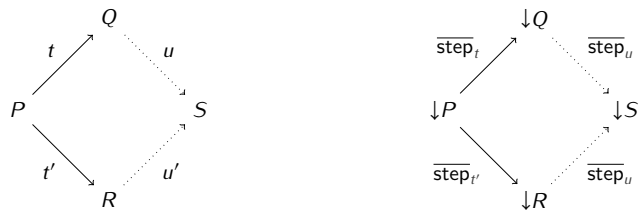
\begin{nscenter}
\begin{minipage}[t]{0.4\columnwidth}
  \begin{nscenter}
\scalebox{0.8}{
\begin{tikzpicture}[node distance=1.5cm, auto]
  \node (P) [node distance=2cm] {
    $P$
  };
  \node (R) [right of=P, above of=P] {
    $Q$
  };
  \node (RPrime) [below of=P, right of=P] {
    $R$
  };
  \node (PPrime) [right of=R, below of=R] {
    $S$
  };
  \draw[->] (P) to node {$t$} (R);
  \draw[->] (P) to node [swap] {$t'$} (RPrime);
  \draw[dotted,->] (R) to node {$u$} (PPrime);
  \draw[dotted,->] (RPrime) to node [swap] {$u'$} (PPrime);
\end{tikzpicture}
}
\end{nscenter}

\end{minipage}%
\begin{minipage}[t]{0.4\columnwidth}
  \begin{nscenter}
\scalebox{0.8}{
\begin{tikzpicture}[node distance=1.5cm, auto]
  \node (P) [node distance=2cm] {
    $\down{P}$
  };
  \node (R) [right of=P, above of=P] {
    $\down{Q}$
  };
  \node (RPrime) [below of=P, right of=P] {
    $\down{R}$
  };
  \node (PPrime) [right of=R, below of=R] {
    $\down{S}$
  };
  \draw[->] (P) to node {$\stepGC{t}$} (R);
  \draw[->] (P) to node [swap] {$\stepGC{t'}$} (RPrime);
  \draw[dotted,->] (R) to node {$\stepGC{u}$} (PPrime);
  \draw[dotted,->] (RPrime) to node [swap] {$\stepGC{u}$} (PPrime);
\end{tikzpicture}
}
\end{nscenter}

\end{minipage}
\end{nscenter}

\vspace{5pt}
\noindent where $\down{P}$ means the lattice of slices of $P$, and
$\stepGC{t}: \down{P} \to \down{Q}$ is a \emph{Galois connection}, a
kind of generalised order isomorphism. An order isomorphism between
posets $A$ and $B$ is a pair of monotone functions $f: A \to B$ and $g:
B \to A$ such that $f \after g = \sub{\id}{B}$ and $g \after f =
\sub{\id}{A}$. Galois connections require only $f \after g \geq
\sub{\id}{B}$ and $g \after f \leq \sub{\id}{A}$ where $\leq$ means the
pointwise order. Galois connections are closed under composition.

The relationship to slicing is that these properties can be unpacked
into statements of sufficiency and minimality: for example $f \after g
\geq \sub{\id}{B}$ means $g$ (backward-slicing) is ``sufficient'' in
that $f$ (forward-slicing) is able to use the result of $g$ to restore
the slicing criterion, and $g \after f \leq \sub{\id}{A}$ means $g$ is
``minimal'' in that it computes the smallest slice with that property.
One can dualise these statements to make similar observations about $f$.

We omit a treatment of structural congruence from our approach, but note
that it slots easily into the framework, generating lattice isomorphisms
in a manner similar to the ``bound braid'' relation $\boundBraid$
discussed in \secref{slicing-concurrent},
\defref{braiding:proc:galois-connection}.

\subsection{Lattices of slices}

The syntax of names, processes and actions is given in
\figref{syntax:basic}. Slices are represented syntactically, via the
$\hole$ notation introduced informally in \secref{introduction}. Our
formalisation employs de Bruijn indices~\cite{debruijn72}, an approach
with well-known strengths and weaknesses compared to other approaches to
names such as higher-order abstract syntax or nominal calculi.

\begin{figure}[h]
\crossrule
\adjustbox{valign=t}{\begin{minipage}[t]{0.5\linewidth}
\begin{syntaxfig}
\mbox{Name}
&
x, y & ::= & 0 \mid 1 \mid \cdots
\\[1mm]
\mbox{Payload}
&
z & ::= & \hole
&
\quad\text{erased}
\\
&&& x
&
\quad\text{retained}
\\[1mm]
\mbox{Action}
&
a
&
::=
&
\hole
&
\quad\text{erased}
\\
&&&
\piInput{x}
&
\quad\text{input}
\\
&&&
\piOutput{x}{z}
&
\quad\text{output}
\\
&&&
\piBoundOutput{x}
&
\quad\text{bound output}
\\
&&&
\piTau
&
\quad\text{silent}
\end{syntaxfig}
\end{minipage}}%
\adjustbox{valign=t}{\begin{minipage}[t]{0.5\linewidth}
\begin{syntaxfig}
\mbox{Process}
&
P, Q, R, S
&
::=
&
\hole
&
\quad\text{erased}
\\
&&&
\piZero
&
\quad\text{inactive}
\\
&&&
\piAction{\piInput{x}}{P}
&
\quad\text{input}
\\
&&&
\piAction{\piOutput{x}{z}}{P}
&
\quad\text{output}
\\
&&&
\piChoice{P}{Q}
&
\quad\text{choice}
\\
&&&
\piPar{P}{Q}
&
\quad\text{parallel}
\\
&&&
\piRestrict{P}
&
\quad\text{restriction}
\\
&&&
\piReplicate{P}
&
\quad\text{replication}
\end{syntaxfig}
\end{minipage}}
\\
\crossrule
\caption{Syntax of names, processes and actions}
\label{fig:syntax:basic}
\end{figure}

\Paragraph{Names}
Only names which occur in the ``payload'' (argument) position of a
message may be erased. The erased name $\hole$ gives rise to a (trivial)
partial order $\leq$ over payloads, namely the partial order containing
precisely $\hole \leq z$ for any $z$. The set of \emph{slices} of $x$ is
written $\down{x}$ and defined to be $\set{z \mid z \leq x}$; because
names are atomic $\down{x}$ is simply the two-element set $\{\hole,
x\}$. The set $\down{x}$ is a finite lattice with meet and join
operations $\meet$ and $\join$, and top and bottom elements $x$ and
$\hole$ respectively. For any lattice, the meet and join are related to
the underlying partial order by $z \leq z' \iff z \join z' = z' \iff z
\meet z' = z$. Lattices are closed under component-wise products,
justifying the notation $\down{(z,z')}$ for $\down{z} \times \down{z'}$.

\Paragraph{Processes}
The $\leq$ relation and $\down{\param}$ operation extend to processes,
via payloads which may be $\hole$, and a special undefined process also
written $\hole$. A slice of $P$ is simply $P$ with some sub-terms
replaced by $\hole$. The relation $\leq$ is the least compatible partial
order which has $\hole$ as least element; all process constructors both
preserve and reflect $\leq$, so we assume an equivalent inductive
definition of $\leq$ when convenient. A process has a closing context
$\Gamma$ enumerating its free variables; in the untyped de Bruijn
setting $\Gamma$ is just a natural number. Often it is convenient to
conflate $\Gamma$ with a set of that cardinality.

\Paragraph{Actions}
An action $a$ labels a transition (\figref{transition} below), and is
either \emph{bound} or \emph{non-bound}. A bound action $b$ is of the
form $\piInput{x}$ or $\piBoundOutput{x}$ and opens a process with
respect to $x$, taking it from $\Gamma$ to $\suc{\Gamma}$. A non-bound
action $c$ is of the form $\piOutput{x}{z}$ or $\piTau$ and preserves
the free variables of the process. The $\leq$ relation and
$\down{\param}$ operation extend to actions via $\hole$ names, plus a
special undefined action also written $\hole$.

\Paragraph{Renamings}
In the lattice setting, a renaming $\rho: \Gamma \to \Gamma'$ is any
function from $\Gamma$ to $\Gamma' \uplus \{\hole\}$; we also allow
$\sigma$ to range over renamings. Renaming application $\ren{\rho}{P}$
is extended with the equation $\ren{\rho}{\hole} = \hole$. The $\leq$
relation and $\down{\param}$ operation apply pointwise.

\vspace{10pt}
\begin{figure}[h]
\begin{smathpar}
\inferrule*
{
}
{
   \piAction{\piInput{x}}{P} \transitionWithoutSmash{\piInput{x}} P
}
\and
\inferrule*
{
}
{
   \piAction{\piOutput{x}{z}}{P} \transitionWithoutSmash{\piOutput{x}{z}} P
}
\and
\inferrule*
{
   P \transitionWithoutSmash{a} R
}
{
   \piChoice{P}{Q} \transitionWithoutSmash{a} R
}
\and
\inferrule*
{
   P \transitionWithoutSmash{c} R
}
{
   \piPar{P}{Q} \transitionWithoutSmash{c} \piPar{R}{Q}
}
\and
\inferrule*[left={$(*)$}]
{
   P \transitionWithoutSmash{b} R
}
{
   \piPar{P}{Q} \transitionWithoutSmash{b} \piPar{R}{\ren{\push{}}{Q}}
}
\and
\inferrule*[left={$(§)$}]
{
   P \transitionWithoutSmash{\piInput{x}} R
   \\
   Q \transitionWithoutSmash{\piOutput{x}{z}} S
}
{
   \piPar{P}{Q} \transitionWithoutSmash{\piTau} \piPar{\ren{(\pop{}{z})}{R}}{S}
}
\and
\inferrule*
{
   P \transitionWithoutSmash{\piOutput{(x + 1)}{0}} R
}
{
   \piRestrict{P} \transitionWithoutSmash{\piBoundOutput{x}} R
}
\and
\inferrule*
{
   P \transitionWithoutSmash{\piInput{x}} R
   \\
   Q \transitionWithoutSmash{\piBoundOutput{x}} S
}
{
   \piPar{P}{Q} \transitionWithoutSmash{\piTau} \piRestrict{(\piPar{R}{S})}
}
\and
\inferrule*[left={$(\dagger)$}]
{
   P \transitionWithoutSmash{\ren{\push{}}{c}} R
}
{
   \piRestrict{P} \transitionWithoutSmash{c} \piRestrict{R}
}
\and
\inferrule*[left={$(\ddagger)$}]
{
   P \transitionWithoutSmash{\ren{\push{}}{b}} R
}
{
   \piRestrict{P} \transitionWithoutSmash{b} \piRestrict{(\ren{\swapR}{R})}
}
\and
\inferrule*
{
   \piPar{P}{\piReplicate{P}} \transitionWithoutSmash{a} R
}
{
   \piReplicate{P} \transitionWithoutSmash{a} R
}
\end{smathpar}
\crossrule
\caption{Labelled transition relation $P \transition{a} R$ (symmetric variants omitted)}
\label{fig:transition}
\end{figure}

\Paragraph{Labelled transition semantics}
The late-style labelled transition semantics is given in
\figref{transition}, and is distinguished only by its adaptation to the
de Bruijn setting. The primary reference for a de Bruijn formulation of
\piCalculus is \cite{hirschkoff99}; the consequences of such an approach
are explored in some depth in \cite{perera16}. One pleasing consequence
of a de Bruijn approach is that the usual side-conditions associated
with transition rules can be operationalised via renamings. We briefly
explain this, along with other uses of renamings in the transition
rules, and refer the interested reader to these earlier works for more
details. \defref{renaming:push-pop-swap} defines the renamings used in
\figref{transition} and \defref{renaming:action} the application
$\ren{\rho}{a}$ of $\rho$ to an action $a$.

\vspace{5pt}
\begin{definition}[$\pushR$, $\popR$, and $\swapR$]
  \label{def:renaming:push-pop-swap}
  \item 
\adjustbox{valign=t}{\begin{minipage}[t]{0.28\linewidth}
{\small
$\begin{array}{llll}
\mathbf{\push{\Gamma}}
&:&
\mathbf{\Gamma \longrightarrow \plus{\Gamma}{1}}
\\
\push{}\;x
&=&
x + 1
\end{array}$
}
\end{minipage}}%
\adjustbox{valign=t}{\begin{minipage}[t]{0.36\linewidth}
{\small
$\begin{array}{llll}
\mathbf{\pop{\Gamma}{z}}
&:&
\mathbf{\plus{\Gamma}{1} \longrightarrow \Gamma}
\\
\pop{}{z}\;0
&=&
z
\\
\pop{}{z}\;(x + 1)
&=&
x
\end{array}$
}
\end{minipage}}%
\adjustbox{valign=t}{\begin{minipage}[t]{0.32\linewidth}
{\small
$\begin{array}{llll}
\mathbf{\swap{\Gamma}}
&:&
\mathbf{\plus{\Gamma}{2} \longrightarrow \plus{\Gamma}{2}}
\\
\swapR\;0
&=&
1
\\
\swapR\;1
&=&
0
\\
\swapR\;(x + 2)
&=&
x + 2
\end{array}$
}
\end{minipage}}

\end{definition}

\begin{definition}[Action renaming]
\label{def:renaming:action}
  Define the following lifting of a renaming to actions.

  \vspace{3pt}
  \begin{nscenter}
{\small
$\begin{array}{llll}
\ren{\param}{}
&&:&
(\Gamma \to \Gamma')
\to
\Action{\Gamma} \to \Action{\Gamma'}
\\
\ren{\rho}{}
&\hole
&=&
\hole
\\
\ren{\rho}{}
&\piInput{x}
&=&
\piInput{\rho x}
\\
\ren{\rho}{}
&\piBoundOutput{x}
&=&
\piBoundOutput{\rho x}
\\
\ren{\rho}{}
&\tau
&=&
\tau
\\
\ren{\rho}{}
&\piOutput{x}{z}
&=&
\smash{\piOutput{\rho x}{\rho z}}
\end{array}$
}
\vspace{3pt}
\end{nscenter}
\crossrule

\end{definition}

\begin{itemize}
\item $\pushR$ occurs in the transition rule which propagates a bound
  action through a parallel composition $\piPar{P}{Q}$ (rule $(*)$ in
  \figref{transition}), and rewires $Q$ so that the name $0$ is
  reserved. The effect is to ensure that the binder being propagated by
  $P$ is not free in $Q$.
\item $\pushR$ also occurs in the rules which propagate an action
  through a $\nu$-binder (rules $(\dagger)$ and $(\ddagger)$), where it
  is applied to the action being propagated using the function defined
  in \defref{renaming:action}. This ensures the action does not mention
  the binder it is propagating through. The use of $\suc{\param}$ in the
  name extrusion rule can be interpreted similarly.
\item $\pop{}{z}$ is used in the event of a successful synchronisation
  (rule $(§)$), and undoes the effect of $\pushR$, substituting the
  communicated name $z$ for index $0$.
\item $\swapR$ occurs in the rule which propagates a bound action
  through a $\nu$-binder (rule $(\dagger)$) and has no counterpart
  outside of the de Bruijn setting. As a propagating binder passes
  through another binder, their relative position in the syntax is
  exchanged, and so to preserve naming $R$ is rewired with a ``braid''
  that swaps $0$ and $1$.
\end{itemize}

\noindent Although its use in the operational semantics is unique to the
de Bruijn setting, $\swapR$ will also play an important role when we
consider the relationship between slices of causally equivalent traces
(\secref{slicing-concurrent} below), where it captures how the relative
position of binders changes between different (but causally equivalent)
interleavings.

\subsection{Galois connections for slicing}

We now compositionally assemble a Galois connection for each component
of execution, starting with renamings, and then proceeding to individual
transitions and entire traces, which relates forward and backward slices
of the initial and terminal state.

\Paragraph{Slicing renamings}
The application $\rho x$ of a renaming to a name, and the lifting
$\ren{\rho}{P}$ of that operation to a process give rise to the Galois
connections defined here.

\begin{definition}[Galois connection for $\rho x$]
Suppose $\rho: \Gamma \to \Gamma'$ and $x \in \Gamma$. Define the
following pair of monotone functions between $\down{(\rho, x)}$ and
$\down{(\rho x)}$.

\adjustbox{valign=t}{\begin{minipage}{0.5\textwidth}
{\small
$\begin{array}{llll}
\renApplF{\rho,x}
&&:&
\down{(\rho, x)} \to \down{(\rho x)}
\\
\renApplF{\rho,x}
&(\sigma,\hole)
&=&
\hole
\\
\renApplF{\rho,x}
&(\sigma,x)
&=&
\sigma x
\end{array}$
}
\end{minipage}}%
\adjustbox{valign=t}{\begin{minipage}{0.5\textwidth}
{\small
$\begin{array}{llll}
\renUnapplF{\rho,x}
&&:&
\down{(\rho x)} \to \down{(\rho, x)}
\\
\renUnapplF{\rho,x}
&z
&=&
(\renMapsTo{\rho}{x}{z}, \renUnapp{\rho}{x}{z})
\end{array}$
}
\end{minipage}}
\\[2mm]
\textit{where}
\adjustbox{valign=t}{\begin{minipage}{0.5\textwidth}
{\small
$\begin{array}{llll}
\renMapsTo{\rho}{x}{\param}
&&:&
\down{(\rho x)} \to \down{\rho}
\\
(\renMapsTo{\rho}{x}{z})&x
&=&
z
\\
(\renMapsTo{\rho}{x}{z})&y
&=&
\hole \quad \text{(if $y \neq x$)}
\end{array}$
}
\end{minipage}}%
\adjustbox{valign=t}{\begin{minipage}{0.5\textwidth}
{\small
$\begin{array}{llll}
\renUnappF{\rho}{x}
&&:&
\down{(\rho x)} \to \down{x}
\\
\renUnappF{\rho}{x}&\hole
&=&
\hole
\\
\renUnappF{\rho}{x}&z
&=&
x \quad \text{(if $z \neq \hole$)}
\end{array}$
}
\end{minipage}}
\\
\crossrule

\end{definition}

\noindent It is convenient to decompose $\renUnapplF{\rho,x}$ into two
components: $\renMapsTo{\rho}{x}{z}$ denotes the least slice of $\rho$
which maps $x$ to $z$, and $\renUnapp{\rho}{x}{z}$ denotes the least
slice of $x$ such that $\rho x = z$.

\begin{lemma}
\label{lem:ren-get:galois-connection}
$(\renApplF{\rho,x}, \renUnapplF{\rho,x})$ is a Galois connection.
\begin{enumerate}
\item $\renApplF{\rho,x} \after \renUnapplF{\rho,x} \geq \sub{\id}{\rho x}$
\item $\renUnapplF{\rho,x} \after \renApplF{\rho,x} \leq \sub{\id}{\rho,x}$
\end{enumerate}
\end{lemma}

\begin{definition}[Galois connection for a renaming $\ren{\rho}{P}$]
\item
Suppose $\rho : \Gamma \to \Gamma'$ and $\Gamma \vdash P$. Define
monotone functions between $\down{(\rho, P)}$ and
$\down{(\ren{\rho}{P})}$ by structural recursion on $\down{P}$, using
the following equations. Here $\hole_{\rho}$ denotes the least slice of $\rho$, namely
the renaming which maps every $x \in \Gamma$ to $\hole$.

\vspace{5pt}
{\small
$\begin{array}{llll}
    \renF{\rho,P}
    &&:&
    \down{(\rho, P)} \to \down{(\ren{\rho}{P})}
    \\
\renF{\rho,P}&(\sigma, \hole)
&=&
\hole
\\
\renF{\rho,\piZero}&(\sigma, \piZero)
&=&
\piZero
\\
\renF{\rho,\piAction{\piInput{x}}{P}}&(\sigma, \piAction{\piInput{x}}{R})
&=&
\piAction{\piInput{x}}{\renF{\suc{\rho},P}\;(\sigma, R)}
\\
\renF{\rho,\piAction{\piOutput{x}{z}}{P}}&(\sigma, \piAction{\piOutput{x}{z'}}{R})
&=&
\piAction{\piOutput{x}{z''}}{\renF{\rho,P}\;(\sigma, R)}
\textit{ where }
z'' = \renAppl{\rho,y}{\sigma,z'}
\\
\renF{\rho,\piChoice{P}{Q}}&(\sigma, \piChoice{R}{S})
&=&
\piChoice{\renF{\rho,P}\;(\sigma, R)}{\renF{\rho,Q}\;(\sigma, S)}
\\
\renF{\rho,\piPar{P}{Q}}&(\sigma,\piPar{R}{S})
&=&
\piPar{\renF{\rho,P}\;(\sigma, R)}{\renF{\rho,Q}\;(\sigma, S)}
\\
\renF{\rho,\piRestrict{P}}&(\sigma, \piRestrict{R})
&=&
\piRestrict{(\renF{\suc{\rho},P}\;(\plus{\sigma}{1},R)}
\\
\renF{\rho,\piReplicate{P}}&(\sigma, \piReplicate{R})
&=&
\piReplicate{(\renF{\rho,P}\;(\sigma,R))}
\\
\\
  \unrenF{\rho,P}
  &&:&
  \down{(\ren{\rho}{P})} \to \down{(\rho, P)}
  \\
\unrenF{\rho,P}&\hole
&=&
(\sub{\hole}{\rho}, \hole)
\\
\unrenF{\rho,\piZero}&\piZero
&=&
(\sub{\hole}{\rho}, \piZero)
\\
\unrenF{\rho,\piAction{\piInput{x}}{P}}&\piAction{\piInput{x}}{R}
&=&
(\rho', \piAction{\piInput{x}}{P'})
\textit{ where }
\unrenF{\suc{\rho},P}\;R = (\suc{\rho'}, P')
\\
\unrenF{\rho,\piAction{\piOutput{x}{z}}{P}}&\piAction{\piOutput{x}{z'}}{R}
&=&
(\rho' \join (\renMapsTo{\rho}{z}{z'}),
\piAction{\piOutput{x}{z''}}{P'})
\textit{ where }
\unrenF{\rho,P}\;R = (\rho', P')
\textit{ and }
z'' = \renUnapp{\rho}{z}{z'}
\\
\unrenF{\rho,\piChoice{P}{Q}}&(\piChoice{R}{S})
&=&
(\rho_1 \join \rho_2, \piChoice{P'}{Q'})
\textit{ where }
\unrenF{\rho,P}\;R = (\rho_1, P')
\text{ and }
\unrenF{\rho,Q}\;S = (\rho_2, Q')
\\
\unrenF{\rho,\piPar{P}{Q}}&(\piPar{R}{S})
&=&
(\rho_1 \join \rho_2, \piPar{P'}{Q'})
\textit{ where }
\unrenF{\rho,P}\;R = (\rho_1, P')
\text{ and }
\unrenF{\rho,Q}\;S = (\rho_2, Q')
\\
\unrenF{\rho,\piRestrict{P}}&\piRestrict{R}
&=&
(\rho', \piRestrict{P'})
\textit{ where }\unrenF{\suc{\rho},P}\;R = (\suc{\rho'} , P')
\\
\unrenF{\rho,\piReplicate{P}}&\piReplicate{R}
&=&
(\rho', \piReplicate{P'})
\textit{ where }\unrenF{\rho,P}\;R = (\rho', P')
\end{array}$
}
\crossrule

\end{definition}

\begin{lemma}
\label{lem:ren:galois-connection}
$(\renF{\rho,P},\unrenF{\rho,P})$ is a Galois connection.
\begin{enumerate}
\item $\renF{\rho,P} \after \unrenF{\rho,P} \geq \sub{\id}{\ren{\rho}{P}}$
\item $\unrenF{\rho,P} \after \renF{\rho,P} \leq \sub{\id}{\rho,P}$
\end{enumerate}
\end{lemma}

\begin{proof}
In each case by induction on $P$, using
\lemref{ren-get:galois-connection} and the invertibility of
$\suc{\param}$.
\end{proof}

\Paragraph{Slicing transitions}
Transitions also lift to the lattice setting, in the form of Galois
connections defined by structural recursion over the proof that $t: P
\transition{a} P'$.
\figrefTwo{transition:fwd-slice}{transition:bwd-slice} define the
forward and backward slicing judgements. We assume a determinising
convention where a rule applies only if no earlier rule applies.

\begin{figure}[t]
\begin{smathpar}
  \inferrule*
  {
    \strut
  }
  {
    \sub{\hole}{P}
    \ftransition{\sub{\hole}{a}}
    \sub{\hole}{P'}
  }
  \and
  \inferrule*
  {
    \strut
  }
  {
    \piAction{\piInput{x}}{\sub{R}{P}}
    \ftransition{\piInput{x}}
    \sub{R}{P}
  }
  \and
  \inferrule*
  {
    \strut
  }
  {
    \piAction{\piOutput{x}{\sub{z'}{z}}}{\sub{R}{P}}
    \ftransition{\piOutput{x}{\sub{z'}{z}}}
    \sub{R}{P}
  }
  \and
  \inferrule*
  {
    \sub{R}{P}
    \ftransition{\smash{\sub{a'}{a}}}
    \sub{R'}{P'}
  }
  {
    \piChoice{\sub{R}{P}}{\sub{S}{Q}}
    \ftransition{\smash{\sub{a'}{a}}}
    \sub{R'}{P'}
  }
  \and
  \inferrule*
  {
    \sub{R}{P}
    \ftransition{\smash{\sub{c'}{c}}}
    \sub{R'}{P'}
  }
  {
    \piPar{\sub{R}{P}}{\sub{S}{Q}}
    \ftransition{\smash{\sub{c'}{c}}}
    \piPar{\sub{R'}{P'}}{\sub{S}{Q}}
  }
  \and
  \inferrule*
  {
    \sub{R}{P}
    \ftransition{\smash{\sub{b'}{b}}}
    \sub{R'}{P'}
  }
  {
    \piPar{\sub{R}{P}}{\sub{S}{Q}}
    \ftransition{\smash{\sub{b'}{b}}}
    \piPar{\sub{R'}{P'}}{\ren{\pushR}{\sub{S}{Q}}}
  }
  \and
  \inferrule*
  {
    \sub{R}{P}
    \ftransition{\piInput{x}}
    \sub{R'}{P'}
    \\
    \sub{S}{Q}
    \ftransition{\piOutput{x}{\smash{\sub{z'}{z}}}}
    \sub{S'}{Q'}
  }
  {
    \piPar{\sub{R}{P}}{\sub{S}{Q}}
    \ftransition{\piTau}
    \piPar{\ren{(\pop{}{\sub{z'}{z}})}{\sub{R'}{P'}}}{\sub{S'}{Q'}}
  }
  \and
  \inferrule*
  {
    \sub{R}{P}
    \ftransition{\sub{\hole}{\piInput{x}}}
    \sub{R'}{P'}
    \\
    \sub{S}{Q}
    \ftransition{\piOutput{x}{\smash{\sub{z'}{z}}}}
    \sub{S'}{Q'}
  }
  {
    \piPar{\sub{R}{P}}{\sub{S}{Q}}
    \ftransition{\sub{\hole}{\piTau}}
    \piPar{\ren{(\pop{}{\sub{z'}{z}})}{\sub{R'}{P'}}}{\sub{S'}{Q'}}
  }
  \and
  \inferrule*
  {
    \sub{R}{P}
    \ftransition{\sub{a}{\piInput{x}}}
    \sub{R'}{P'}
    \\
    \sub{S}{Q}
    \ftransition{\sub{\hole}{\smash{\piOutput{x}{z}}}}
    \sub{S'}{Q'}
  }
  {
    \piPar{\sub{R}{P}}{\sub{S}{Q}}
    \ftransition{\sub{\hole}{\piTau}}
    \piPar{\ren{(\pop{}{\sub{\hole}{z}})}{\sub{R'}{P'}}}{\sub{S'}{Q'}}
  }
  \and
  \inferrule*
  {
    \sub{R}{P}
    \ftransition{\piOutput{(x + 1)}{0}}
    \sub{R'}{P'}
  }
  {
    \piRestrict{\sub{R}{P}}
    \ftransition{\piBoundOutput{x}}
    \sub{R'}{P'}
  }
  \and
  \inferrule*
  {
    \sub{R}{P}
    \ftransition{\sub{a}{\smash{\piOutput{(x + 1)}{0}}}}
    \sub{R'}{P'}
  }
  {
    \piRestrict{\sub{R}{P}}
    \ftransition{\sub{\hole}{\piBoundOutput{x}}}
    \sub{R'}{P'}
  }
  \and
  \inferrule*
  {
    \sub{R}{P}
    \ftransition{\piInput{x}}
    \sub{R'}{P'}
    \\
    \sub{S}{Q}
    \ftransition{\piBoundOutput{x}}
    \sub{S'}{Q'}
  }
  {
    \piPar{\sub{R}{P}}{\sub{S}{Q}}
    \ftransition{\piTau}
    \piRestrict{(\piPar{\sub{R'}{P'}}{\sub{S'}{Q'}})}
  }
  \and
  \inferrule*
  {
    \sub{R}{P}
    \ftransition{\sub{a}{\piInput{x}}}
    \sub{R'}{P'}
    \\
    \sub{S}{Q}
    \ftransition{\sub{a'}{\piBoundOutput{x}}}
    \sub{S'}{Q'}
  }
  {
    \piPar{\sub{R}{P}}{\sub{S}{Q}}
    \ftransition{\sub{\hole}{\piTau}}
    \piRestrict{(\piPar{\sub{R'}{P'}}{\sub{S'}{Q'}})}
  }
  \and
  \inferrule*
  {
    \sub{R}{P}
    \ftransition{\ren{\pushR}{\smash{\sub{c'}{c}}}}
    \sub{R'}{P'}
  }
  {
    \piRestrict{\sub{R}{P}}
    \ftransition{\smash{\sub{c'}{c}}}
    \piRestrict{\sub{R'}{P'}}
  }
  \and
  \inferrule*
  {
    \sub{R}{P}
    \ftransition{\ren{\pushR}{\smash{\sub{b'}{b}}}}
    \sub{R'}{P'}
  }
  {
    \piRestrict{\sub{R}{P}}
    \ftransition{\smash{\sub{b'}{b}}}
    \piRestrict{(\ren{\swapR}{\sub{R'}{P'}})}
  }
  \and
  \inferrule*
  {
    \piPar{\sub{R}{P}}{\piReplicate{\sub{R}{P}}}
    \ftransition{\smash{\sub{a'}{a}}}
    \sub{R'}{P'}
  }
  {
    \piReplicate{\sub{R}{P}}
    \ftransition{\smash{\sub{a'}{a}}}
    \sub{R'}{P'}
  }
\end{smathpar}
\crossrule
\caption{Forward slicing judgement $\sub{R}{P}
  \ftransition{\smash{\sub{a'}{a}}} \sub{R'}{P'}$}
\label{fig:transition:fwd-slice}
\end{figure}

\begin{figure}[t]
\begin{smathpar}
  \inferrule*
  {
    \strut
  }
  {
    \sub{\hole}{P}
    \rtransition{\sub{\hole}{a}}
    \sub{\hole}{P'}
  }
  \and
  \inferrule*
  {
    \strut
  }
  {
    \piAction{\piInput{x}}{\sub{R}{P}}
    \rtransition{\sub{a}{\piInput{x}}}
    \sub{R}{P}
  }
  \and
  \inferrule*
  {
    \strut
  }
  {
    \piAction{\piOutput{x}{\sub{z'}{z}}}{\sub{R}{P}}
    \rtransition{\piOutput{x}{\sub{z'}{z}}}
    \sub{R}{P}
  }
  \and
  \inferrule*
  {
    \strut
  }
  {
    \piAction{\piOutput{x}{\sub{\hole}{z}}}{\sub{R}{P}}
    \rtransition{\sub{\hole}{\piOutput{x}{z}}}
    \sub{R}{P}
  }
  \and
  \inferrule*
  {
    \sub{R'}{P}
    \rtransition{\smash{\sub{a'}{a}}}
    \sub{R}{P'}
  }
  {
    \piChoice{\sub{R'}{P}}{\sub{\hole}{Q}}
    \rtransition{\smash{\sub{a'}{a}}}
    \sub{R}{P'}
  }
  \and
  \inferrule*
  {
    \sub{R'}{P}
    \rtransition{\smash{\sub{c'}{c}}}
    \sub{R}{P'}
  }
  {
    \piPar{\sub{R'}{P}}{\sub{S}{Q}}
    \rtransition{\smash{\sub{c'}{c}}}
    \piPar{\sub{R}{P'}}{\sub{S}{Q}}
  }
  \and
  \inferrule*
  {
    \sub{R'}{P}
    \rtransition{\smash{\sub{c'}{c}}}
    \sub{\hole}{P'}
  }
  {
    \piPar{\sub{R'}{P}}{\sub{\hole}{Q}}
    \rtransition{\smash{\sub{c'}{c}}}
    \sub{\hole}{\piPar{P'}{Q}}
  }
  \and
  \inferrule*
  {
    \sub{R'}{P}
    \rtransition{\smash{\sub{b'}{b}}}
    \sub{R}{P'}
  }
  {
    \piPar{\sub{R'}{P}}{\sub{S}{Q}}
    \rtransition{\smash{\sub{b'}{b}}}
    \piPar{\sub{R}{P'}}{\ren{\sub{\rho}{\pushR}}{\sub{S}{Q}}}
  }
  \and
  \inferrule*
  {
    \sub{R'}{P}
    \rtransition{b}
    \sub{\hole}{P'}
  }
  {
    \piPar{\sub{R'}{P}}{\sub{\hole}{Q}}
    \rtransition{b}
    \sub{\hole}{\piPar{P'}{\ren{\pushR}{Q}}}
  }
  \and
  \inferrule*[right={$\rho 0 = z'$}]
  {
    \sub{R'}{P}
    \rtransition{\piInput{x}}
    \sub{R}{P'}
    \\
    \sub{S'}{Q}
    \rtransition{\piOutput{x}{\sub{z'}{z}}}
    \sub{S}{Q'}
  }
  {
    \piPar{\sub{R'}{P}}{\sub{S'}{Q}}
    \rtransition{\piTau}
    \piPar{\ren{\sub{\rho}{\pop{}{z}}}{\sub{R}{P'}}}{\sub{S}{Q'}}
  }
  \and
  \inferrule*[right={$\rho 0 = z$}]
  {
    \sub{R'}{P}
    \rtransition{\sub{\hole}{\piInput{x}}}
    \sub{R}{P'}
    \\
    \sub{S'}{Q}
    \rtransition{\piOutput{x}{z}}
    \sub{S}{Q'}
  }
  {
    \piPar{\sub{R'}{P}}{\sub{S'}{Q}}
    \rtransition{\sub{\hole}{\piTau}}
    \piPar{\ren{\sub{\rho}{\pop{}{z}}}{\sub{R}{P'}}}{\sub{S}{Q'}}
  }
  \and
  \inferrule*[right={$\rho 0 = \hole$}]
  {
    \sub{R'}{P}
    \rtransition{\sub{\hole}{\piInput{x}}}
    \sub{R}{P'}
    \\
    \sub{S'}{Q}
    \rtransition{\sub{\hole}{\piOutput{x}{z}}}
    \sub{S}{Q'}
  }
  {
    \piPar{\sub{R'}{P}}{\sub{S'}{Q}}
    \rtransition{\sub{\hole}{\piTau}}
    \piPar{\ren{\sub{\rho}{\pop{}{z}}}{\sub{R}{P'}}}{\sub{S}{Q'}}
  }
  \and
  \inferrule*
  {
    \sub{R}{P}
    \rtransition{\piInput{x}}
    \sub{\hole}{P'}
    \\
    \sub{S}{Q}
    \rtransition{\piOutput{x}{\smash{\sub{\hole}{z}}}}
    \sub{\hole}{Q'}
  }
  {
    \piPar{\sub{R}{P}}{\sub{S}{Q}}
    \rtransition{\piTau}
    \sub{\hole}{\piPar{\ren{(\pop{}{z})}{P'}}{Q'}}
  }
  \and
  \inferrule*
  {
    \sub{R'}{P}
    \rtransition{\piOutput{(x + 1)}{0}}
    \sub{R}{P'}
  }
  {
    \piRestrict{\sub{R'}{P}}
    \rtransition{\sub{a}{\piBoundOutput{x}}}
    \sub{R}{P'}
  }
  \and
  \inferrule*
  {
    \sub{R'}{P}
    \rtransition{\piInput{x}}
    \sub{R}{P'}
    \\
    \sub{S'}{Q}
    \rtransition{\piBoundOutput{x}}
    \sub{S}{Q'}
  }
  {
    \piPar{\sub{R'}{P}}{\sub{S'}{Q}}
    \rtransition{\piTau}
    \piRestrict{(\piPar{\sub{R}{P'}}{\sub{S}{Q'}})}
  }
  \and
  \inferrule*
  {
    \sub{R'}{P}
    \rtransition{\sub{\hole}{\piInput{x}}}
    \sub{R}{P'}
    \\
    \sub{S'}{Q}
    \rtransition{\sub{\hole}{\piBoundOutput{x}}}
    \sub{S}{Q'}
  }
  {
    \piPar{\sub{R'}{P}}{\sub{S'}{Q}}
    \rtransition{\sub{\hole}{\piTau}}
    \piRestrict{(\piPar{\sub{R}{P'}}{\sub{S}{Q'}})}
  }
  \and
  \inferrule*
  {
    \sub{R}{P}
    \rtransition{\piInput{x}}
    \sub{\hole}{P'}
    \\
    \sub{S}{Q}
    \rtransition{\piBoundOutput{x}}
    \sub{\hole}{Q'}
  }
  {
    \piPar{\sub{R}{P}}{\sub{S}{Q}}
    \rtransition{\piTau}
    \piRestrict{\sub{\hole}{\piPar{P'}{Q'}}}
  }
  \and
  \inferrule*
  {
    \sub{R}{P}
    \rtransition{\sub{\hole}{\piInput{x}}}
    \sub{\hole}{P'}
    \\
    \sub{S}{Q}
    \rtransition{\sub{\hole}{\piBoundOutput{x}}}
    \sub{\hole}{Q'}
  }
  {
    \piPar{\sub{R}{P}}{\sub{S}{Q}}
    \rtransition{\sub{\hole}{\piTau}}
    \piRestrict{\sub{\hole}{\piPar{P'}{Q'}}}
  }
  \and
  \inferrule*
  {
    \sub{R}{P}
    \rtransition{\piInput{x}}
    \sub{\hole}{P'}
    \\
    \sub{S}{Q}
    \rtransition{\piBoundOutput{x}}
    \sub{\hole}{Q'}
  }
  {
    \piPar{\sub{R}{P}}{\sub{S}{Q}}
    \rtransition{\piTau}
    \sub{\hole}{\piRestrict{(\piPar{P'}{Q'})}}
  }
  \and
  \inferrule*
  {
    \sub{R'}{P}
    \rtransition{\ren{\pushR}{c'}}
    \sub{R}{P'}
  }
  {
    \piRestrict{\sub{R'}{P}}
    \rtransition{\smash{\sub{c'}{c}}}
    \piRestrict{\sub{R}{P'}}
  }
  \and
  \inferrule*
  {
    \sub{R'}{P}
    \rtransition{\ren{\pushR}{c'}}
    \sub{\hole}{P'}
  }
  {
    \piRestrict{\sub{R'}{P}}
    \rtransition{\smash{\sub{c'}{c}}}
    \sub{\hole}{\piRestrict{P'}}
  }
  \and
  \inferrule*
  {
    \sub{R'}{P}
    \rtransition{\ren{\pushR}{b'}}
    \sub{R}{P'}
  }
  {
    \piRestrict{\sub{R'}{P}}
    \rtransition{\smash{\sub{b'}{b}}}
    \piRestrict{(\ren{\sub{\rho}{\swapR}}{\sub{R}{P'}})}
  }
  \and
  \inferrule*
  {
    \sub{R'}{P}
    \rtransition{\ren{\pushR}{b}}
    \sub{\hole}{P'}
  }
  {
    \piRestrict{\sub{R'}{P}}
    \rtransition{b}
    \sub{\hole}{\piRestrict{(\ren{\swapR}{P'})}}
  }
  \and
  \inferrule*
  {
    \piPar{\sub{R'}{P}}{\sub{R^{\twoPrime}}{\piReplicate{P}}}
    \rtransition{\smash{\sub{a'}{a}}}
    \sub{R}{P'}
  }
  {
    \sub{(\piReplicate{R'} \join R^{\twoPrime})}{\piReplicate{P}}
    \rtransition{\smash{\sub{a'}{a}}}
    \sub{R}{P'}
  }
\end{smathpar}
\crossrule
\caption{Backward slicing judgement $\smash{\sub{R'}{P}
    \rtransition{\smash{\sub{a'}{a}}} \sub{R}{P'}}$}
\label{fig:transition:bwd-slice}
\end{figure}

The judgement $\sub{R}{P} \ftransition{\smash{\sub{a'}{a}}}
\sub{R'}{P'}$ asserts that there is a ``replay'' transition from $R \leq
P$ to $(a',R') \leq (a,P)$, with $R$ the input and $(a', R')$ the
output. The judgement $\smash{\sub{R'}{P}
  \rtransition{\sub{\smash{a'}}{a}} \sub{R}{P'}}$ asserts that there is
a ``rewind'' transition from $(a', R) \leq (a, P')$ to $R' \leq P$, with
$(a', R)$ the input and $R'$ the output. When writing $\sub{R}{P}$ where
$R \leq P$ we exploit the preservation and reflection of $\leq$ by all
constructors, for example writing
$\piRestrict{(\piPar{\sub{R}{P}}{\sub{S}{Q}})}$ for
$\sub{\piRestrict{(\piPar{R}{S})}}{\piRestrict{(\piPar{P}{Q})}}$.

For backward slicing, we permit the renaming
application operator $\ren{}{}$ to be used in a pattern-matching form,
indicating a use of the lower adjoint $\unrenF{}$: given a renaming
application $\ren{\rho}{P}$, the pattern $\ren{\sigma}{P'}$ matches any
slice $R$ of $\ren{\rho}{P}$ such that $\unrenF{\rho,P}(R) =
(\sigma,P')$.

\begin{definition}[Galois connection for a transition]
\label{def:galois-connection:step-unstep}
Suppose $t: P \transition{a} P'$. Define the following pair of monotone
functions between $\down{P}$ to $\down{(a, P')}$.

\adjustbox{valign=t}{\begin{minipage}{0.5\textwidth}
{\small
\[\begin{array}{llll}
\stepF{t}
&&:&
\down{P} \to \down{(a, P')}
\\
\stepF{t}
&
R
&=&
(a',R')\textit{ where }\sub{R}{P} \ftransition{\sub{a'}{\smash{a}}} \sub{R'}{P'}
\end{array}\]}
\end{minipage}}%
\adjustbox{valign=t}{\begin{minipage}{0.5\textwidth}
{\small
\[\begin{array}{llll}
\unstepF{t}
&&:& \down{(a, P')} \to \down{P}
\\
\unstepF{t}
&
(R,a')
&=&
R'\textit{ where }\sub{R'}{P} \rtransition{\sub{a'}{\smash{a}}} \sub{R}{P'}
\end{array}\]}
\end{minipage}}
\crossrule
\end{definition}

\noindent We omit the proofs that these equations indeed define total,
deterministic, monotone relations.

\begin{theorem}[$(\stepF{t},\unstepF{t})$ is a Galois connection]
\item
\begin{enumerate}
\item $\stepF{t} \after \unstepF{t} \geq \sub{\id}{a,P'}$
\item $\unstepF{t} \after \stepF{t} \leq \sub{\id}{P}$
\end{enumerate}
\end{theorem}

\begin{proof}
By induction on $t: P \transition{a'} P'$, using
\lemref{ren:galois-connection} for the cases involving renaming.
\end{proof}

\Paragraph{Slicing traces} Finally we extend
slicing to entire runs of a \piCalculus program. A sequence of
transitions $\vec{t}$ is called a \emph{trace}; the empty trace at $P$
is written $\sub{\nil}{P}$, and the composition of a transition $t: P
\transition{a} R$ and trace $\vec{t}: R \transition{\vec{a}} S$ is
written $t \cons \vec{t}: P \transition{a \cons \vec{a}} S$ where
actions are composable whenever their source and target contexts match.

\begin{definition}[Galois connection for a trace]
Suppose $\vec{t}: P \transition{\vec{a}} P'$. Define the following pair
of monotone functions between $\down{P}$ and $\down{P'}$, using variants
of $\stepF{t}$ and $\unstepF{t}$ which discard the action slice (going
forward) and which use $\hole$ as the action slice (going backward).

\crossrule
\\
\begin{minipage}{0.5\textwidth}
{\small
$\begin{array}{lll}
  \fwdF{\vec{t}}
  &:&
  \down{P} \to \down{P'}
  \\
  \fwd{\sub{\nil}{P}}{}
  &=&
  \sub{\id}{\down{P}}
  \\
  \fwd{t \cons \vec{t}}{\hole}
  &=&
  \hole
  \\
  \fwd{t \cons \vec{t}}{R}
  &=&
  \fwd{\vec{t}}{(\stepNoAction{t}{R})}
  \qquad
  (R \neq \hole)
  \\
  \\
  \mathbf{\stepNoActionF{t}}
  &:&
  \mathbf{\down{P} \to \down{P'}}
  \\
  \stepNoAction{t}{R}
  &=&
  R'
  \textit{ where }
  \step{t}{R} = (a',R')
\end{array}$
}
\end{minipage}%
\begin{minipage}{0.5\textwidth}
{\small
$\begin{array}{lll}
  \bwdF{\vec{t}}
  &:&
  \down{P'} \to \down{P}
  \\
  \bwd{\sub{\nil}{P'}}{}
  &=&
  \sub{\id}{\down{P'}}
  \\
  \bwd{t \cons \vec{t}}{\hole}
  &=&
  \hole
  \\
  \bwd{t \cons \vec{t}}{R}
  &=&
  \unstepNoAction{t}{(\bwd{\vec{t}}{R})}
  \qquad
  (R \neq \hole)
  \\
  \\
  \unstepNoActionF{t}
  &:&
  \down{P'} \to \down{P}
  \\
  \unstepNoAction{t}{R'}
  &=&
  \unstep{t}{(\hole, R')}
\end{array}$
}
\end{minipage}
\\
\crossrule

\end{definition}

\noindent At the empty trace $\sub{\nil}{P}$ the Galois connection is
simply the identity on $\down{P}$. Otherwise, we recurse into the
structure of the trace $t \cons \vec{t}$, composing the Galois
connection for the single transition $t$ with the Galois connection for
the tail of the computation $\vec{t}$.

\begin{theorem}[$(\fwdF{\vec{t}},\bwdF{\vec{t}})$ is a Galois connection]
\item
\begin{enumerate}
\item $\fwdF{\vec{t}} \after \bwdF{\vec{t}} \geq \sub{\id}{P'}$
\item $\bwdF{\vec{t}} \after \fwdF{\vec{t}} \leq \sub{\id}{P}$
\end{enumerate}
\end{theorem}

\noindent Note that the trace used to define forward and backward
slicing for a computation is not an auxiliary data structure recording
the computation, such as a redex trail or memory, but simply the proof
term witnessing $P \transition{\vec{a}} P'$.

\section{Slicing and causal equivalence}
\label{sec:slicing-concurrent}

In this section, we show that when dynamic slicing a \piCalculus
program, slicing with respect to any causally equivalent execution
yields essentially the same slice. ``Essentially the same'' here means
modulo lattice isomorphism. In other words slicing discards precisely
the same information regardless of which interleaving is chosen to do
the slicing.

\Paragraph{Proof-relevant causal equivalence}
Causally equivalent computations are generated by transitions which
share a start state, but which are independent. Following L\'evy
\cite{levy80}, we call such transitions \emph{concurrent}, written $t
\concur t'$. We illustrate this idea, and the non-trivial relationship
that it induces between terminal states, by way of example. For the full
definition of concurrency for \piCalculus, we refer the interested
reader to \cite{perera16} or to the Agda
definition\footnote{\url{https://github.com/rolyp/proof-relevant-pi/blob/master/Transition/Concur.agda}}.
For the sake of familiarity the example uses regular names instead of de
Bruijn indices.

\Paragraph{Example} Consider the process $P_0 \eqdef
\piRestrictN{yz}{\piPar{(\piAction{\piOutput{x}{y}}{P})}{\piAction{\piOutput{x}{z}}{Q}}}$
for some unspecified processes $P$ and $Q$. This process can take
\emph{two} transitions, which we will call $t$ and $t'$. Transition $t:
P_0 \transition{\piBoundOutputN{x}{y}} P_1$ extrudes $y$ on the channel
$x$:

\vspace{-8pt}
\begin{smathpar}
P_0 \transition{\piBoundOutputN{x}{y}}
\piRestrictN{z}{\piPar{P}{\piAction{\piOutput{x}{z}}{Q}}} \eqdef P_1
\end{smathpar}
whereas transition $t': P_0 \transition{\piBoundOutputN{x}{z}} P_1'$
extrudes $z$, also on the channel $x$:

\vspace{-8pt}
\begin{smathpar}
P_0 \transition{\piBoundOutputN{x}{z}}
\piRestrictN{y}{\piPar{(\piAction{\piOutput{x}{y}}{P})}{Q}} \eqdef P_1'
\end{smathpar}
In both cases the output actions are bound, representing the extruding
binder. Moreover, $t$ and $t'$ are \emph{concurrent}, written $t \concur
t'$, meaning they can be executed in either order. Having taken $t$, one
can \emph{mutatis mutandis} take $t'$, and vice versa. Concurrency is an
irreflexive and symmetric relation defined over transitions which are
\emph{coinitial} (have the same source state).

The qualification is needed because $t'$ will need to be adjusted to
operate on the target state of $t$, if $t$ is the transition which
happens first. If $t'$ happens first then $t$ will need to be adjusted
to operate on the target state of $t'$. The adjusted version of $t'$ is
called the \emph{residual} of $t'$ after $t$, and is written
$\residual{t'}{t}$. In this case $\residual{t'}{t}$ can still extrude
$z$:

\vspace{-10pt}
\begin{smathpar}
P_1 = \piRestrictN{z}{\piPar{P}{\piAction{\piOutput{x}{z}}{Q}}}
\transition{\piBoundOutputN{x}{z}} \piPar{P}{Q} \eqdef P_0'
\end{smathpar}
whereas the residual $\residual{t}{t'}$ can still extrude $y$:
\begin{smathpar}
P_1' = \piRestrictN{y}{\piPar{(\piAction{\piOutput{x}{y}}{P})}{Q}}
\transition{\piBoundOutputN{x}{y}}
\piPar{P}{Q} = P_0'
\end{smathpar}

\noindent The independence of $t$ and $t'$ is confirmed by the fact that
$t \cons \residual{t'}{t}$ and $t' \cons \residual{t}{t'}$ are
\emph{cofinal} (share a target state), as shown on the left below.
  \begin{nscenter}
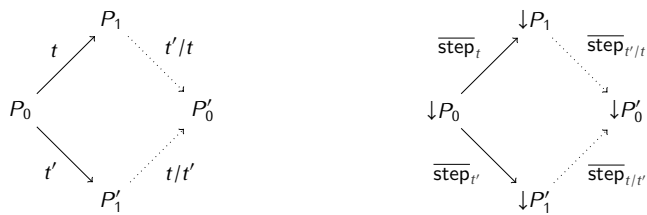
  \begin{minipage}[t]{0.4\columnwidth}
    \begin{nscenter}
\scalebox{0.8}{
\begin{tikzpicture}[node distance=1.5cm, auto]
  \node (P) [node distance=2cm] {
    $P_0$
  };
  \node (R) [right of=P, above of=P] {
    $P_1$
  };
  \node (RPrime) [below of=P, right of=P] {
    $P_1'$
  };
  \node (PPrime) [right of=R, below of=R] {
    $P_0'$
  };
  \draw[->] (P) to node {$t$} (R);
  \draw[->] (P) to node [swap] {$t'$} (RPrime);
  \draw[dotted,->] (R) to node {$\residual{t'}{t}$} (PPrime);
  \draw[dotted,->] (RPrime) to node [swap] {$\residual{t}{t'}$} (PPrime);
\end{tikzpicture}
}
\end{nscenter}

  \end{minipage}%
  \begin{minipage}[t]{0.4\columnwidth}
    \begin{nscenter}
\scalebox{0.8}{
\begin{tikzpicture}[node distance=1.5cm, auto]
  \node (P) [node distance=2cm] {
    $\down{P_0}$
  };
  \node (R) [right of=P, above of=P] {
    $\down{P_1}$
  };
  \node (RPrime) [below of=P, right of=P] {
    $\down{P_1'}$
  };
  \node (PPrime) [right of=R, below of=R] {
    $\down{P_0'}$
  };
  \draw[->] (P) to node {$\stepGC{t}$} (R);
  \draw[->] (P) to node [swap] {$\stepGC{t'}$} (RPrime);
  \draw[dotted,->] (R) to node {$\stepGC{\residual{t'}{t}}$} (PPrime);
  \draw[dotted,->] (RPrime) to node [swap] {$\stepGC{\residual{t}{t'}}$} (PPrime);
\end{tikzpicture}
}
\end{nscenter}

  \end{minipage}
  \end{nscenter}

\vspace{5pt}
\noindent We say that the traces $\smash{\vec{t} \eqdef t \cons
  \residual{t'}{t}}$ and $\smash{\vec{u} \eqdef t' \cons
  \residual{t}{t'}}$ are \emph{causally equivalent}, written $\vec{t}
\permEq \vec{u}$. The commutativity of the right-hand square
(\thmref{residual:slicing:pentagon} below) means the two interleavings
are also equivalent for slicing purposes. Here $\stepGC{t}$ denotes the
Galois connection $(\stepF{t},\unstepF{t})$.

However \cite{perera16}, which formalised causal equivalence for
\piCalculus, showed that causally equivalence traces do not always reach
exactly the same state, but only the same state up to some permutation
of the binders in the resulting processes. This will become clear if we
consider another process $Q_0 \eqdef
\piPar{(\piAction{\piInputN{x}{y'}}{R})}{\piAction{\piInputN{x}{z'}}{S}}$
able to synchronise with both of the extrusions raised by $P_0$ and
consider the two different ways that $\piPar{P_0}{Q_0}$ can evolve.

First note that $Q_0$ can also take two independent transitions: $u: Q_0
\transition{\piInputN{x}{y'}} \piPar{R}{\piAction{\piInputN{x}{z'}}{S}}
\eqdef Q_1$ inputs on $x$ and binds the received name to $y'$; and $u':
Q_0 \transition{\piInputN{x}{z'}}
\piPar{(\piAction{\piInputN{x}{y'}}{R})}{S} \eqdef Q_1'$ also inputs on
$x$ and binds the received name to $z'$. (Assume $z$ is not free in the
left-hand side of $Q_0$ and that $y$ is not free in the right-hand
side.) The respective residuals $Q_1 =
\piPar{R}{\piAction{\piInputN{x}{z'}}{S}} \transition{\piInputN{x}{z'}}
\piPar{R}{S} \eqdef Q_0'$ and $Q_1' =
\piPar{(\piAction{\piInputN{x}{y'}}{R})}{S}
\transition{\piInputN{x}{y'}} \piPar{R}{S} = Q_0'$ again converge on the
same state $Q_0'$, leading to a diamond for $Q_0$ similar to the one for
$P_0$ above.

The subtlety arises when we put $P_0$ and $Q_0$ into parallel
composition, since now we have two concurrent synchronisation
possibilities. For clarity we give the derivations, which we call $s$
and $s'$:

\vspace{-5pt}
\begin{smathpar}
  \inferrule*
  {
    t: P_0 \transition{\piBoundOutputN{x}{y}} P_1
    \\
    u: Q_0 \transition{\piInputN{x}{y'}} Q_1
  }
  {
    s: \piPar{P_0}{Q_0} \transition{\tau} \piRestrictN{y}{\piPar{P_1}{\subst{Q_1}{y}{y'}}}
  }
  \and
  \inferrule*
  {
    t': P_0 \transition{\piBoundOutputN{x}{z}} P_1'
    \\
    u': Q_0 \transition{\piInputN{x}{z'}} Q_1'
  }
  {
    s': \piPar{P_0}{Q_0} \transition{\tau} \piRestrictN{z}{\piPar{P_1'}{\subst{Q_1'}{z}{z'}}}
  }
\end{smathpar}

\noindent The labelled transition system is closed under renamings; thus
the residual $\residual{u'}{u}$ has an image in the renaming
$\subst{\param}{y}{y'}$, and $\residual{u}{u'}$ has an image in the
renaming $\subst{\param}{z}{z'}$, allowing us to derive composite
residual $\residual{s'}{s}$:

\vspace{-5pt}
\begin{smathpar}
  \inferrule*
  {
    \residual{t'}{t}: P_1 \transition{\piBoundOutputN{x}{z}} P_0'
    \\
    \inferrule*
    {
      \residual{u'}{u}: Q_1 \transition{\piInputN{x}{z'}} Q_0'
    }
    {
      \subst{(\residual{u'}{u})}{y}{y'}: \subst{Q_1}{y}{y'} \transition{\piInputN{x}{z'}} \subst{Q_0'}{y}{y'}
    }
  }
  {
    \residual{s'}{s}: \piRestrictN{y}{\piPar{P_1}{\subst{Q_1}{y}{y'}}} \transition{\tau}
    \piRestrictN{yz}{\piPar{P_0'}{\subst{\subst{Q_0'}{y}{y'}}{z}{z'}}}
  }
\end{smathpar}

\noindent By similar reasoning we can derive $\residual{s}{s'}$:

\begin{center}
{\small$\residual{s}{s'}: \piRestrictN{z}{\piPar{P_1'}{\subst{Q_1'}{y}{y'}}} \transition{\tau}
    \piRestrictN{zy}{\piPar{P_0'}{\subst{\subst{Q_0'}{z}{z'}}{y}{y'}}}$}
\end{center}

\noindent By side-conditions on the transition rules the renamings
$\subst{\param}{y}{y'}$ and $\subst{\param}{z}{z'}$ commute and so
$\subst{\subst{Q_0'}{y}{y'}}{z}{z'} \eqdef Q_0^\twoPrime =
\subst{\subst{Q_0'}{z}{z'}}{y}{y'}$. However, the positions of binders
$y$ and $z$ are transposed in the terminal states of $\residual{s'}{s}$
and $\residual{s}{s'}$. Instead of the usual diamond shape, we have the
pentagon on the left below, where $\phi$ is a \emph{braid} representing
the transposition of the binders. Lifted to slices, $\phi$ becomes the
unique isomorphism $\braidGC{\phi}$ relating slices of the terminal
states, as shown in the commutative diagram on the right:

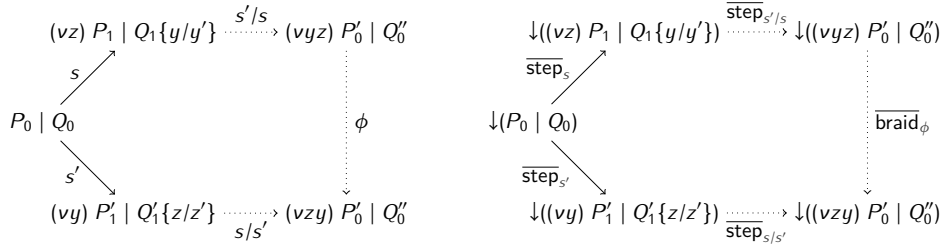
\begin{nscenter}
\begin{minipage}[t]{0.45\columnwidth}
  \scalebox{0.8}{
\begin{tikzpicture}[node distance=1.5cm, auto, baseline=(P.base)]
  \node (P) [node distance=2cm] {
    $\piPar{P_0}{Q_0}$
  };
  \node (R) [right of=P, above of=P] {
    $\piRestrictN{z}{\piPar{P_1}{\subst{Q_1}{y}{y'}}}$
  };
  \node (RPrime) [below of=P, right of=P] {
    $\piRestrictN{y}{\piPar{P_1'}{\subst{Q_1'}{z}{z'}}}$
  };
  \node (Q) [node distance=3.5cm, right of=R] {
    $\piRestrictN{yz}{\piPar{P_0'}{Q_0''}}$
  };
  \node (QPrime) [node distance=3.5cm, right of=RPrime] {
    $\piRestrictN{zy}{\piPar{P_0'}{Q_0''}}$
  };
  \draw[->] (P) to node [yshift=-1ex] {$s$} (R);
  \draw[->] (P) to node [yshift=1ex,swap] {$s'$} (RPrime);
  \draw[dotted,->] (R) to node {$\residual{s'}{s}$} (Q);
  \draw[dotted,->] (RPrime) to node [swap] {$\residual{s}{s'}$} (QPrime);
  \draw[dotted,->] (Q) to node {$\phi$} (QPrime);
\end{tikzpicture}
}

\end{minipage}%
\begin{minipage}[t]{0.45\columnwidth}
\scalebox{0.8}{
\begin{tikzpicture}[node distance=1.5cm, auto, baseline=(P.base)]
  \node (P) [node distance=2cm] {
    $\down{(\piPar{P_0}{Q_0})}$
  };
  \node (R) [right of=P, above of=P] {
    $\down{(\piRestrictN{z}{\piPar{P_1}{\subst{Q_1}{y}{y'}}})}$
  };
  \node (RPrime) [below of=P, right of=P] {
    $\down{(\piRestrictN{y}{\piPar{P_1'}{\subst{Q_1'}{z}{z'}}})}$
  };
  \node (Q) [node distance=4cm, right of=R] {
    $\down{(\piRestrictN{yz}{\piPar{P_0'}{Q_0''}})}$
  };
  \node (QPrime) [node distance=4cm, right of=RPrime] {
    $\down{(\piRestrictN{zy}{\piPar{P_0'}{Q_0''}})}$
  };
  \draw[->] (P) to node [yshift=-1ex] {$\stepGC{s}$} (R);
  \draw[->] (P) to node [yshift=1ex,swap] {$\stepGC{s'}$} (RPrime);
  \draw[dotted,->] (R) to node {$\stepGC{\residual{s'}{s}}$} (Q);
  \draw[dotted,->] (RPrime) to node [swap] {$\stepGC{\residual{s}{s'}}$} (QPrime);
  \draw[dotted,->] (Q) to node {$\braidGC{\phi}$} (QPrime);
\end{tikzpicture}
}

\end{minipage}
\end{nscenter}

In the de Bruijn setting, a braid like $\phi$ does not relate two
processes of the form $\piRestrictN{yz}{R}$ and $\piRestrictN{yz}{R}$
but rather two processes of the form $\piRestrict{\piRestrict{R}}$ and
$\piRestrict{\piRestrict{(\ren{\swapR}{R})}}$: the transposition of the
(nameless) binders is represented by the transposition of the roles of
indices $0$ and $1$ in the body of the innermost binder.

\begin{definition}[Bound braid $P \boundBraid R$]
  \label{def:bound-braid}
  Inductively define the symmetric relation $P \boundBraid R$ using the
  rules below.

  \vspace{-8pt}
  \begin{smathpar}
\inferrule*[left={\textnormal{$\congRestrictSwap{P}$}},right={\textnormal{$P = \ren{\swapR}{P'}$}}]
{
   \strut
}
{
   \piRestrict{\piRestrict{P}}
   \boundBraid
   \piRestrict{\piRestrict{P'}}
}
\and
\inferrule*[left={$\piChoice{\param}{Q}$}]
{
   P \boundBraid R
}
{
   \piChoice{P}{Q}
   \boundBraid
   \piChoice{R}{Q}
}
\and
\inferrule*[left={$\piChoice{P}{\param}$}]
{
   Q \boundBraid S
}
{
   \piChoice{P}{Q}
   \boundBraid
   \piChoice{P}{S}
}
\and
\inferrule*[left={$\piPar{\param}{Q}$}]
{
   P \boundBraid R
}
{
   \piPar{P}{Q}
   \boundBraid
   \piPar{R}{Q}
}
\and
\inferrule*[left={$\piPar{P}{\param}$}]
{
   Q \boundBraid S
}
{
   \piPar{P}{Q}
   \boundBraid
   \piPar{P}{S}
}
\and
\inferrule*[left={$\piRestrict{\param}$}]
{
   P \boundBraid R
}
{
   \piRestrict{P}
   \boundBraid
   \piRestrict{R}
}
\and
\inferrule*[left={$\piReplicate{\param}$}]
{
   P \boundBraid R
}
{
   \piReplicate{P}
   \boundBraid
   \piReplicate{R}
}
\end{smathpar}
\crossrule

\end{definition}

\noindent Following \cite{perera16}, we adopt a compact term-like
notation for $\boundBraid$ proofs, using the rule names which occur to
the left of each rule in \defref{bound-braid}. For the extrusion example
above, $\phi$ (in de Bruijn indices notation) would be a leaf case of
the form $\congRestrictSwap{\piPar{\param}{\param}}$.

\begin{definition}[Lattice isomorphism for bound braid]
\label{def:braiding:proc:galois-connection}
  Suppose $\phi: Q \boundBraid Q'$. Define the following pair of
  monotone functions between $\down{Q}$ and $\down{Q'}$ by structural
  recursion on $\phi$.

  \vspace{3pt}
\adjustbox{valign=t}{\begin{minipage}{0.5\textwidth}
{\small
$\begin{array}{llll}
  \braidF{\phi}
  &&:&
  \down{Q} \to \down{Q'}
  \\
  \braidF{\congRestrictSwap{P}}
  &(\piRestrict{\piRestrict{R}})
  &=&
  \piRestrict{\piRestrict{(\renF{\swapR,P}(R))}}
  \\
  \braidF{\piChoice{\phi}{S}}
  &(\piChoice{R}{S})
  &=&
  \piChoice{\braid{\phi}{R}}{S}
  \\
  \braidF{\piChoice{R}{\psi}}
  &(\piChoice{R}{S})
  &=&
  \piChoice{R}{\braid{\psi}{S}}
  \\
  \braidF{\piPar{\phi}{S}}
  &(\piPar{R}{S})
  &=&
  \piPar{\braid{\phi}{R}}{S}
  \\
  \braidF{\piPar{R}{\psi}}
  &(\piPar{R}{S})
  &=&
  \piPar{R}{\braid{\psi}{S}}
  \\
  \braidF{\piRestrict{\phi}}
  &(\piRestrict{R})
  &=&
  \piRestrict{(\braid{\phi}{R})}
  \\
  \braidF{\piReplicate{\phi}}
  &(\piReplicate{R})
  &=&
  \piReplicate{(\braid{\phi}{R})}
\end{array}$
}
\end{minipage}}%
\adjustbox{valign=t}{\begin{minipage}{0.5\textwidth}
{\small
$\begin{array}{llll}
  \unbraidF{\phi}
  &&:&
  \down{Q'} \to \down{Q}
  \\
  \unbraidF{\congRestrictSwap{P}}&{(\piRestrict{\piRestrict{R}})}
  &=&
  \piRestrict{\piRestrict{(\renF{\swapR,P}(R))}}
  \\
  \unbraidF{\piChoice{\phi}{S}}&{(\piChoice{R}{S})}
  &=&
  \piChoice{\unbraid{\phi}{R}}{S}
  \\
  \unbraidF{\piChoice{R}{\psi}}&{(\piChoice{R}{S})}
  &=&
  \piChoice{R}{\unbraid{\psi}{S}}
  \\
  \unbraidF{\piPar{\phi}{S}}&{(\piPar{R}{S})}
  &=&
  \piPar{\unbraid{\phi}{R}}{S}
  \\
  \unbraidF{\piPar{R}{\psi}}&{(\piPar{R}{S})}
  &=&
  \piPar{R}{\unbraid{\psi}{S}}
  \\
  \unbraidF{\piRestrict{\phi}}&{(\piRestrict{R})}
  &=&
  \piRestrict{(\unbraid{\phi}{R})}
  \\
  \unbraidF{\piReplicate{\phi}}&{(\piReplicate{R})}
  &=&
  \piReplicate{(\unbraid{\phi}{R})}
\end{array}$
}
\end{minipage}}
\\
\crossrule

\end{definition}

\begin{lemma}
  \item
  \begin{enumerate}
    \item $\braidF{\phi} \after \unbraidF{\phi} = \sub{\id}{\down{Q'}}$
    \item $\unbraidF{\phi} \after \braidF{\phi} = \sub{\id}{\down{Q}}$
  \end{enumerate}
\end{lemma}

\begin{proof}
  Induction on $\phi$. In the base case use the idempotence of $\swapR$
  lifted to lattices.
\end{proof}

\begin{definition}[Lattice isomorphism for cofinality map]
  Suppose $t \concur t'$ with $\target{\residual{t'}{t}} = Q$ and
  $\target{\residual{t}{t'}} = Q'$. By Theorem 1 of \cite{perera16},
  there exists a unique $\braiding{t,t'}$ witnessing $Q = Q'$, $Q
  \freeBraid Q'$ or $Q \boundBraid Q'$. Define the following pair of
  monotone functions between $\down{Q}$ and $\down{Q'}$.

  \vspace{3pt}
\begin{minipage}{0.5\textwidth}
{\small
$\begin{array}{llll}
  \mapF{\braiding{t,t'}}
  &:&
  \down{Q} \to \down{Q'}
  \\
  \mapF{Q = Q'}
  &=&
  \sub{\id}{\down{Q}}
  \\
  \mapF{Q \freeBraid Q'}
  &=&
  \renF{\swapR,Q}
  \\
  \mapF{\phi: Q \boundBraid Q'}
  &=&
  \braidF{\phi}
\end{array}$
}
\end{minipage}%
\begin{minipage}{0.5\textwidth}
{\small
$\begin{array}{llll}
  \unmapF{\braiding{t,t'}}
  &:&
  \down{Q'} \to \down{Q}
  \\
  \unmapF{Q = Q'}
  &=&
  \sub{\id}{\down{Q}}
  \\
  \unmapF{Q \freeBraid Q'}
  &=&
  \unrenF{\swapR,Q}
  \\
  \unmapF{\phi: Q \boundBraid Q'}
  &=&
  \unbraidF{\phi}
\end{array}$
}
\end{minipage}
\\
\crossrule

\end{definition}

\begin{lemma}
  \item
  \begin{enumerate}
    \item $\mapF{\braiding{t,t'}} \after \unmapF{\braiding{t,t'}} = \sub{\id}{\down{Q'}}$
    \item $\unmapF{\braiding{t,t'}} \after \mapF{\braiding{t,t'}} = \sub{\id}{\down{Q}}$
  \end{enumerate}
\end{lemma}

\begin{theorem}
\label{thm:residual:slicing:pentagon}
  Suppose $t \concur t'$ as on the left. Then the pentagon on the right
  commutes.

  \begin{nscenter}
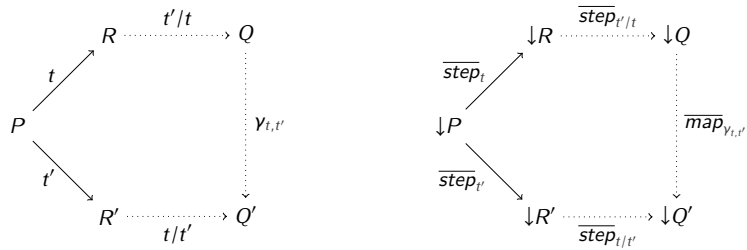
  \begin{minipage}[t]{0.4\columnwidth}
    \scalebox{0.8}{
\begin{tikzpicture}[node distance=1.5cm, auto, baseline=(P.base)]
  \node (P) [node distance=2cm] {
    $P$
  };
  \node (R) [right of=P, above of=P] {
    $R$
  };
  \node (RPrime) [below of=P, right of=P] {
    $R'$
  };
  \node (Q) [node distance=2.25cm, right of=R] {
    $Q$
  };
  \node (QPrime) [node distance=2.25cm, right of=RPrime] {
    $Q'$
  };
  \draw[->] (P) to node [yshift=-1ex] {$t$} (R);
  \draw[->] (P) to node [yshift=1ex,swap] {$t'$} (RPrime);
  \draw[dotted,->] (R) to node {$\residual{t'}{t}$} (Q);
  \draw[dotted,->] (RPrime) to node [swap] {$\residual{t}{t'}$} (QPrime);
  \draw[dotted,->] (Q) to node {$\braiding{t,t'}$} (QPrime);
\end{tikzpicture}
}

  \end{minipage}%
  \begin{minipage}[t]{0.4\columnwidth}
  \scalebox{0.8}{
\begin{tikzpicture}[node distance=1.5cm, auto, baseline=(P.base)]
  \node (P) [node distance=2cm] {
    $\down{P}$
  };
  \node (R) [right of=P, above of=P] {
    $\down{R}$
  };
  \node (RPrime) [below of=P, right of=P] {
    $\down{R'}$
  };
  \node (Q) [node distance=2.25cm, right of=R] {
    $\down{Q}$
  };
  \node (QPrime) [node distance=2.25cm, right of=RPrime] {
    $\down{Q'}$
  };
  \draw[->] (P) to node [yshift=-1ex] {$\stepGC{t}$} (R);
  \draw[->] (P) to node [yshift=1ex,swap] {$\stepGC{t'}$} (RPrime);
  \draw[dotted,->] (R) to node {$\stepGC{\residual{t'}{t}}$} (Q);
  \draw[dotted,->] (RPrime) to node [swap] {$\stepGC{\residual{t}{t'}}$} (QPrime);
  \draw[dotted,->] (Q) to node {$\mapGC{\braiding{t,t'}}$} (QPrime);
\end{tikzpicture}
}

  \end{minipage}
  \end{nscenter}
\end{theorem}

\Paragraph{Lattice isomorphism for arbitrary causal equivalence}
Concurrent transitions $t \concur t'$ induce an ``atom'' of causal
equivalence, ${t \cons \residual{t'}{t}} \permEq {t' \cons
  \residual{t}{t'}}$. The full relation is generated by closing under
the trace constructors (for horizontal composition) and transitivity
(for vertical composition). In \cite{perera16} this yields a composite
form of cofinality map $\braiding{\alpha}$ where $\alpha: \vec{t}
\permEq \vec{u}$ is an arbitrary causal equivalence. We omit further
discussion for reasons of space, but note that $\braiding{\alpha}$ is
built by composing and translating (by contexts) atomic cofinality maps,
and so gives rise, by composition of isomorphisms, to a lattice
isomorphism between $\down{\target{\vec{t}}}$ and
$\down{\target{\vec{u}}}$.

\section{Related work}
\label{sec:related-work}

\Paragraph{Reversible process calculi} Reversible process calculi
have recently been used for speculative execution, debugging,
transactions, and distributed protocols that require backtracking. A key
challenge is to permit backwards execution to leverage concurrency
whilst ensuring causal consistency. In contrast to our work, reversible
calculi focus on mechanisms for reversibility, such as the thread-local
memories used by Danos and Krivine's reversible CCS \cite{danos04a},
Lanese \etal's $\rho\pi$ \cite{lanese10}, and Cristescu \etal's
reversible \piCalculus \cite{cristescu13}. We intentionally remain
agnostic about implementation strategy, whilst providing a formal
guarantee that causally consistent rewind and replay are a suitable
foundation for any implementation.

\Paragraph{Concurrent program slicing} An early example of concurrent dynamic slicing is
Duesterwald \etal, who consider a language with synchronous
message-passing \cite{duesterwald92}. They give a notion of correctness
with respect to a slicing criterion, but find that computing least
slices is undecidable, in contrast to our slices which are extremal by
construction. Following Cheng~\cite{cheng93}, most subsequent work has
recast dynamic slicing as a dependency-graph reachability problem; our
approach is to slice with respect to a particular interleaving, but show
how to derive the slice corresponding to any execution with the same
dependency structure.

Goswami and Mall consider shared-memory concurrency \cite{goswami00},
and Mohapatra \etal tackle slicing for concurrent Java
\cite{manandhar04}, but both present only algorithms, with no formal
guarantees. Tallam \etal develop an approach based on dependency graphs,
but again offer only algorithms and empirical results \cite{tallam08}.
Moreover most prior work restricts the slicing criteria to the (entire)
values of particular variables, rather than arbitrary parts of
configurations.

\Paragraph{Provenance and slicing} Our interest in
slicing arises in part due to connections with provenance, and recent
applications of provenance to security~\cite{acar13}. Others have also
considered provenance models in concurrency calculi, including Souliah
\etal~\cite{souilah09} and Dezani-Ciancaglini
\etal~\cite{dezani-ciancaglini12}. Further study is needed to relate our
approach to provenance and security.

\section{Conclusion}
\label{sec:conclusion}

The main contribution of this paper is to extend our previous approach
to slicing based on Galois connections to \piCalculus, and show that the
resulting notion of slice is invariant under causal equivalence. For
this latter step, we build on a prior formalisation of causal
equivalence for \piCalculus~\cite{perera16}. Although de Bruijn indices
significantly complicate the resulting definitions, the formalism is
readily implemented in Agda. This paper provides a foundation for future
development of rigorous provenance tracing or dynamic slicing techniques
for practical concurrent programs, which we plan to investigate in
future work.

\subparagraph*{Acknowledgements.}

Perera and Cheney were supported by the Air Force Office of Scientific
Research, Air Force Material Command, USAF, under grant number
FA8655-13-1-3006. The U.S. Government and University of Edinburgh are
authorized to reproduce and distribute reprints for their purposes
notwithstanding any copyright notation thereon. Perera was also
supported by UK EPSRC project EP/K034413/1. Umut Acar helped with
problem formulation and an earlier approach. Vít Šefl provided valuable
Agda technical support. Our anonymous reviewers provided useful
comments.

\bibliographystyle{plainurl}

\begin{thebibliography}{}

\end{thebibliography}


\begin{thebibliography}{10}

\bibitem{acar13}
Umut~A. Acar, Amal Ahmed, James Cheney, and Roly Perera.
\newblock A core calculus for provenance.
\newblock {\em Journal of Computer Security}, 21:919--969, 2013.
\newblock Full version of a POST 2012 paper.

\bibitem{cheng93}
Jingde Cheng.
\newblock Slicing concurrent programs: A graph-theoretical approach.
\newblock In {\em Automated and Algorithmic Debugging}, number 749 in LNCS,
  pages 223--240. Springer-Verlag, 1993.

\bibitem{cristescu13}
Ioana Cristescu, Jean Krivine, and Daniele Varacca.
\newblock A compositional semantics for the reversible pi-calculus.
\newblock pages 388--397, June 2013.

\bibitem{danos04a}
Vincent Danos and Jean Krivine.
\newblock Reversible communicating systems.
\newblock In {\em Concurrency Theory, 15th International Conference, CONCUR
  '04}, pages 292--307. Springer, 2004.

\bibitem{debruijn72}
N.G. de~Bruijn.
\newblock Lambda-calculus notation with nameless dummies: a tool for automatic
  formula manipulation with application to the {Church}-{Rosser} theorem.
\newblock {\em Indagationes Mathematicae}, 34(5):381--392, 1972.

\bibitem{dezani-ciancaglini12}
M.~Dezani{-}Ciancaglini, R.~Horne, and V.~Sassone.
\newblock Tracing where and who provenance in linked data: {A} calculus.
\newblock {\em Theoretical Computer Science}, 464:113--129, 2012.

\bibitem{duesterwald92}
Evelyn Duesterwald, Rajiv Gupta, and Mary~Lou Soffa.
\newblock Distributed slicing and partial re-execution for distributed
  programs.
\newblock In {\em Proceedings of the 5th International Workshop on Languages
  and Compilers for Parallel Computing}, pages 497--511. Springer, 1993.

\bibitem{goswami00}
D.~Goswami and R.~Mall.
\newblock Dynamic slicing of concurrent programs.
\newblock In {\em High Performance Computing -- HiPC 2000}, volume 1970, pages
  15--26. Springer, 2000.

\bibitem{hirschkoff99}
Daniel Hirschkoff.
\newblock Handling substitutions explicitly in the pi-calculus.
\newblock In {\em 2nd International Workshop on Explicit Substitutions: Theory
  and Applications to Programs and Proofs}, 1999.

\bibitem{lanese10}
Ivan Lanese, Claudio~Antares Mezzina, and Jean-Bernard Stefani.
\newblock Reversing higher-order~$\pi$.
\newblock In {\em Concurrency Theory, 21st International Conference, CONCUR
  '10}, pages 478--493. Springer, 2010.

\bibitem{levy80}
Jean-Jacques L\'evy.
\newblock Optimal reductions in the lambda-calculus.
\newblock In J.~P. Seldin and J.~R. Hindley, editors, {\em To H. B. Curry:
  Essays in Combinatory Logic, Lambda Calculus and Formalism}, pages 159--191.
  Academic Press, New York, NY, USA, 1980.

\bibitem{milner99}
Robin Milner.
\newblock {\em Communicating and mobile systems: the $\pi$ calculus}.
\newblock Cambridge University Press, Cambridge, UK, 1999.

\bibitem{manandhar04}
D.P. Mohapatra, Rajib Mall, and Rajeev Kumar.
\newblock An efficient technique for dynamic slicing of concurrent {Java}
  programs.
\newblock In {\em Applied Computing}, volume 3285, pages 255--262. Springer,
  2004.

\bibitem{perera12a}
Roly Perera, Umut~A. Acar, James Cheney, and Paul~Blain Levy.
\newblock Functional programs that explain their work.
\newblock In {\em 17th ACM SIGPLAN International Conference on Functional
  Programming}, ICFP '12, pages 365--376. ACM, 2012.

\bibitem{perera16}
Roly Perera and James Cheney.
\newblock Proof-relevant pi-calculus, 2016.
\newblock Submitted to \textit{Mathematical Structures in Computer Science}.
  \url{http://arxiv.org/abs/1604.04575}.

\bibitem{souilah09}
Issam Souilah, Adrian Francalanza, and Vladimiro Sassone.
\newblock A formal model of provenance in distributed systems.
\newblock In {\em TAPP 2009}, Berkeley, CA, USA, 2009. USENIX Association.

\bibitem{tallam08}
Sriraman Tallam, Chen Tian, and Rajiv Gupta.
\newblock Dynamic slicing of multithreaded programs for race detection.
\newblock In {\em 24th IEEE International Conference on Software Maintenance
  (ICSM 2008)}, pages 97--106. IEEE, 2008.

\bibitem{weiser81}
Mark Weiser.
\newblock Program slicing.
\newblock In {\em Proceedings of the 5th International Conference on Software
  Engineering}, ICSE '81, pages 439--449, Piscataway, NJ, USA, 1981. IEEE
  Press.

\end{thebibliography}

\pagebreak
\begin{appendix}

\section{Agda module structure}
\label{app:module-structure}

\figref{modules} summarises the module structure of the repository
\ConcurrentSlicing, which contains the Agda formalisation. The module
structure of the auxiliary repositories is described in \cite{perera16}.
All repositories can be found at the URL \url{https://github.com/rolyp}.
\\ \begin{figure}[h]
{\small
\noindent \begin{tabular}{lp{10cm}}
\text{\emph{Auxiliary repositories}}
\\
\ttt{agda-stdlib-ext 0.0.3}
& Extensions to Agda library
\\
\ttt{proof-relevant-pi 0.3}
& Concurrent transitions, residuals and causal equivalence
\\
\\
\text{\emph{Core modules}}
\\
\tt{Action.Lattice}
& Action slices $a' \in \down{a}$
\\
\tt{Action.Concur.Lattice}
& Action residual, lifted to slices
\\
\tt{Action.Ren.Lattice}
& Action renaming, lifing to slices
\\
\tt{Braiding.Proc.Lattice}
& Bound braids, lifted to slices via $\braidF{\phi}$ and $\unbraidF{\phi}$
\\
\tt{ConcurrentSlicing}
& Include everything; compile to build project
\\
\tt{ConcurrentSlicingCommon}
& Common imports from standard library
\\
\tt{Example}
& Milner's scheduler example
\\
\tt{Example.Helper}
& Utility functions for examples
\\
\tt{Lattice}
& Lattice typeclass
\\
\tt{Lattice.Product}
& Component-wise product of lattices
\\
\tt{Name.Lattice}
& Name slices $y \in \down{x}$
\\
\tt{Proc.Lattice}
& Process slices $P' \in \down{P}$
\\
\tt{Proc.Ren.Lattice}
& Process renaming, lifted to slices via $\renF{\rho,P}$ and $\unrenF{\rho,P}$
\\
\tt{Ren.Lattice}
& Renaming slices $\sigma \in \down{\rho}$ and application to slices ($\renApplF{\rho,x}$ and $\renUnapplF{\rho,x}$)
\\
\tt{Ren.Lattice.Properties}
& Additional properties relating to renaming slices
\\
\tt{Transition.Lattice}
& Slicing functions $\stepF{t}$ and $\unstepF{gt}$
\\
\tt{Transition.Ren.Lattice}
& Renaming of transitions, lifted to lattices
\\
\\
\tt{Transition.Concur.Cofinal.Lattice}
& Braidings $\braiding{t,t'}$ lifted to slices
\\
\tt{Transition.Seq.Lattice}
& Slicing functions $\fwdF{\vec{t}}$ and $\bwdF{\vec{t}}$
\\
\\
\text{\emph{Common sub-modules}}
\\
\tt{.GaloisConnection}
& Galois connection between lattices defined in parent module
\\ 
\\
\end{tabular}
}
\crossrule
\caption{\ConcurrentSlicing module overview, release \ttt{0.1}}
\label{fig:modules}
\end{figure}

\end{appendix}

\end{document}